\documentclass[reprint,amsmath,amssymb,aps,pre]{revtex4-2} 

\usepackage[utf8]{inputenc}
\usepackage{times,amsmath,amsfonts,amssymb,latexsym,mathtools}
\usepackage{graphicx}
\usepackage{subfigure}
\usepackage[normalem]{ulem}
\usepackage{physics, outlines, bm}
\usepackage{amsthm}
\usepackage{hyperref}
\hypersetup{colorlinks = true,allcolors = blue }
\newcommand{\eye}{\mathbb{I}} %\pmb
\newcommand{\bseq}{\begin{subequations}}
\newcommand{\eseq}{\end{subequations}}
\newcommand{\map}{\mathcal{M}}
\newcommand{\torder}{\mathcal T}

\newcommand{\ov}[1]{\overline{#1}}
\newcommand\h{{\cal H}}

\usepackage{amsthm}

\usepackage[capitalize,nameinlink]{cleveref}

\newcommand{\bsplit}{\begin{split}}
\newcommand{\esplit}{\end{split}}

\begin{document}
\title{Universally-Charging Protocols for Quantum Batteries: A No-Go Theorem}
\author{Pratik Sathe}
\email{psathe@lanl.gov}
\author{Francesco Caravelli}
\email{caravelli@lanl.gov}
\affiliation{T-Division (T-4), Los Alamos National Laboratory, Los Alamos (NM), 87544 USA} %Bikini Atoll road, 

\date{\today}

\begin{abstract}
    The effectiveness of a quantum battery relies on a robust charging process, yet these are often sensitive to initial state of the battery.
    We introduce the concept of a universally-charging (UC) protocol, defined as one that either increases or maintains the average battery energy for all initial states, without ever decreasing it.
    We show that UC protocols are impossible for closed quantum batteries, thus necessitating interactions with auxilliary quantum systems.
    To that end, we prove a no-go theorem which prohibits UC protocols for closed quantum batteries with finite-dimensional Hilbert spaces.
    Leveraging a no-go theorem for topological quantum walks, we argue that even for infinite-dimensional Hilbert spaces, while unitary UC operators exist, they cannot be generated by physically reasonable Hamiltonian protocols.
    However, regardless of the dimension, non-unitary UC protocols can be achieved in open quantum batteries.
    To illustrate this, we present a general model with a control qubit, whose state interpolates between universal-charging and universal-discharging protocols.
\end{abstract}

\maketitle
\setcounter{footnote}{0} 

\emph{Introduction.}---
Quantum technologies promise to outperform classical technologies by harnessing uniquely phenomena such as entanglement, superposition and coherence.
Substantial evidence points to the plausibility of quantum advantage in computation~\cite{DaleyPractical2022,GrumblingQuantum2019}, sensing~\cite{KantsepolskyExploring2023,DegenQuantum2017}, and cryptography~\cite{RennerQuantum2023}. %AruteQuantum2019
Significant interest has also emerged in leveraging quantum correlations to enhance the performance of quantum thermodynamic devices, such as quantum heat engines and quantum batteries~\cite{AlickiEntanglement2013, BhattacharjeeQuantum2021, Auffves2022}.
Early foundational work demonstrated enhanced work extraction from entangled states in quantum batteries, demonstrating that entanglement can significantly boost the charging power of quantum batteries~\cite{campaioli}.

Several studies have investigated the role of quantum coherence and entanglement in optimizing the charging process for quantum batteries~\cite{AlickiEntanglement2013, BhattacharjeeQuantum2021,FerraroHighPower2018}, leading to faster charging times and higher energy storage efficiency~\cite{Salvia2023,Seah2021}.
Other research has explored the scaling of quantum battery performance with system size, revealing that collective quantum effects can enhance the energy extraction process in ways that classical systems cannot replicate~\cite{PoliniPRB2018,PoliniPRL2019,PoliniPRB2019}. 
There are also various investigations of the dynamics of quantum batteries \cite{Ferraro2018,Le2018,Rossini2020,Shaghaghi2022,topquantbat}, which can lead to rapid energy redistribution and optimized charging processes~\cite{andolinasyk}. 
Analyses of the robustness of quantum batteries against environmental noise and decoherence have also established bounds on their operational performance~\cite{Binder2015,LewensteinBatteries18,Bakhshinezhad2024,Caravelli2021,Shi2022}.
In particular, it has been shown that global operations are required to obtain an advantage in charging speedups~\cite{Gyhm2022}.
Quantum batteries have also been experimentally realized~\cite{Quach2022,Yu2024,Joshi2022}.

Classical, electrochemical batteries are charged by a single protocol, regardless of the battery's initial state.
Similarly, given an initial state of a closed quantum battery, it is always possible (with some exceptions) to design a unitary protocol that charges the battery.
While designing and optimizing charging protocols that demonstrate a quantum advantage are promising and important objectives, the dependence of such protocols on the state of the battery, which can itself be deviate from the desired state, remains less explored.
In this letter, along these lines, we explore and highlight a subtle but key limitation of a fully coherent quantum battery due to the unitarity of the processes.

With the goal of defining robust charging protocols, we seek those that increase or maintain the battery (average) energy but never reduce it for any initial state.
We call such protocols universally-charging (UC).
We show that UC protocols can be achieved only for open quantum batteries, thus necessitating an auxiliary system or bath.
To that end, we show that for a closed quantum battery, it is impossible to construct a UC protocol.
In some sense, this is similar to the no-cloning theorem in quantum mechanics, which states that it is impossible to copy an unknown quantum state~\cite{Wootters1982}. 
Furthermore, we show that for any given unitary protocol, the amount of energy charged, averaged across all possible initial states is identically zero. 
Therefore, a charging protocol that is optimized for a certain initial state, might be ineffective or may even discharge the battery when the initial state deviates significantly from the intended one.

While similar observations in specific cases have been made, to our knowledge, this phenomenon has not been appreciated in its generality.
For example, as noted in Ref.~\cite{Santos2019} and the references therein, a charging interaction has to be switched off when the fully charged state is reached, to avoid spontaneous discharging.

This theorem is directly applicable only to quantum batteries with finite-dimensional Hilbert spaces.
It can be overcome if the battery has an infinite-dimensional Hilbert space, or if the battery is an open quantum system, resulting in a non-unitary charging protocol.
In the former case, we show that if the battery Hamiltonian has a single-sided infinite energy ladder (similar to the spectrum of a harmonic oscillator), unitary UC protocols still do not exist.
However, UC unitaries, such as the energy-shift operator~\cite{AbergCatalytic2014}, do exist for double-sided infinite energy ladders.
By interpreting the energy-shift operator as a quantum walk~\cite{KadianQuantum2021} in the energy space, we show that it has a non-zero topological index~\cite{GrossIndex2012}, and consequently, cannot be generated by any Hamiltonian that is local~\cite{LiuClassification2023} in the energy basis of the battery Hamiltonian, and therefore is not physically realizable.
We argue that this property is more general, and all UC unitary protocols are thus physically unrealizable on double-sided energy ladders.

We demonstrate the existence of \textit{non-unitary} UC protocols by presenting a recipe for constructing them.
Taking inspiration from models of topological quantum walks~\cite{KitagawaExploring2010} and chiral Floquet topological insulators~\cite{LiuChiral2018}, we present a UC protocol by coupling any finite-dimensional quantum battery to an auxiliary qubit. 
In the limit of a infinite-dimensional quantum battery Hilbert space, the auxiliary qubit becomes a catalyst, and the protocol becomes unitary, without conflict with the topological obstruction mentioned above.

\emph{Setup.}---
We consider closed as well as open quantum batteries.
Following standard convention~\cite{CampaioliColloquium2024}, the energy of the battery is computed with respect to the battery Hamiltonian $H_B$, so that the (average) battery energy $E = \Tr [H_B \rho_B]$, with $\rho_B$ denoting the battery density matrix.
For an open quantum battery, the total Hilbert space $\h_{\text{total}} = \h_B \otimes \h_{E}$, where $B$ and $E$ denote the battery and environment respectively. 
In this case, $\rho_B = \Tr_E \rho_{\text{total}}$, the partial trace of the full density matrix.

A charging protocol may be regarded as a mapping of battery density matrices to themselves. 
They can therefore be expressed as CPTP (completely positive trace-preserving) quantum channels~\cite{Breuer2007}.
The (average) amount of energy charged by a protocol $\map$ for an initial state $\rho_B$ is $\Delta_\map E = \Tr (H_B \{ \map [\rho_B] - \rho_B\})$. 
We will drop the subscript $\map$ for $\Delta E$ when clear from the context.
We consider cases where $\dim \h_B$ is finite as well as infinite. 
We assume that the energy eigenvalues of $H_B$ are countable when infinite in number. 

We define a universally-charging (UC) protocol $\map$, as one for which $\Delta_\map E(\rho_B) \geq 0 \ \forall \rho_B$ and $\Delta_\map E(\rho_B) > 0$ for at least one $\rho_B$. 
(The latter condition is imposed to exclude trivial maps such as the identity map.)
Universally-discharging (UD) protocols can be defined similarly.

A UC map $\map$ is unitary if there exists  a unitary operator $U$, such that $\map[\rho] = U\rho U^\dagger$ for any $\rho$.
Clearly, if $U$ is UC unitary operator, then its conjugate, $U^\dagger$, defines a UD unitary protocol.

\emph{Closed batteries with finite-dimensional Hilbert spaces.}--
Let us consider closed quantum batteries, so that charging protocols are necessarily unitary.
First, we note that for (almost) every density matrix, there exists a unitary map that increases its energy.
This can be seen easily by noting that the problem of charging with respect to $H_B$ is equivalent to discharging with respect to $-H_B$. 
Thus, the only states that cannot be charged using a unitary protocol are those whose ergotropy~\cite{Nieuwenhuizen} with respect to $-H_B$ is zero.
It follows from Refs.~\cite{LenardThermodynamical1978,PuszPassive1978,Salviadistribution2021} that the only states that cannot be charged by any unitary operation are diagonal in the eigenbasis of $H_B$ with eigenvalues non-decreasing in energy. 

Flipping the question around, we ask whether a protocol can charge every state, and find the answer to be negative:
There does not exist any UC unitary protocol for a battery with a finite dimensional Hilbert space.
We call this the \textit{no-charging theorem}.

We provide two proofs for this statement. 
For the first one, we start by assuming that there exists a unitary $U$ such that $\Delta E (\rho_s) \coloneqq \Tr [H_B \{ U \rho_S U^\dagger - \rho_S\}] \geq 0$ for all $\rho_S$.
For simplicity, we assume that the energy eigenvalues of $H_B$, denoted $E_1 < E_2 < \dotsc < E_N$ are all non-degenerate. 
(The proof can be straightforwardly modified for the case with degeneracies.) 
Let us denote the corresponding eigenvectors by $\ket{1}, \ket{2}, \dotsc, \ket{N}$.
Thus, $H_B = \sum_{n=1}^N E_n \ket n \bra n$.
Expressing $U$ in the eigenbasis of $H_B$, we have $U  = \sum_{m,n=1}^N c_{m,n} \ket{m}\bra n$ with $\sum_{m=1}^N \abs{c_{m,n}}^2 = 1 \ \forall \ n$ due to unitarity.
Choosing $\rho_S = \ket N \bra N$, we have $\Delta E(\ket N \bra N) \geq 0$.
This implies that 
\bseq
\begin{align}
        \Tr [H_0 U \ket{N}\bra{N} U^\dagger ] & \geq E_N, \\
        \implies \sum_{m} \abs{c_{m,N}}^2 E_m &\geq E_N, \\
        \implies \abs{c_{m,N}} &= \delta_{m,N}.
\end{align}
\eseq
Therefore, $U \ket N = e^{i\theta_N} \ket N$ for some $\theta_N$.
Furthermore, $c_{N,N-1} = \bra N U \ket{N-1} = e^{i\theta_N} \bra N \ket{N-1} = 0$.
Therefore, $U$ maps the subspace spanned by $\{ \ket{1},\dotsc, \ket{N-1}\}$ back to itself.
Thus, using $\Delta E (\ket {N-1} \bra{N-1}) \geq 0$, we get $U \ket{N-1} = e^{i\phi_{N-1}} \ket{N-1}$.
Following the same steps as before, we conclude that $U \ket{j} = e^{i\phi_j} \ket j $ for all $j$.
We immediately conclude that $\Delta E (\rho_S) =0$ for all $\rho_S$, and consequently, $U$ does not (positively) charge any single density matrix.
It follows that $U$ cannot define a UC protocol.

We note that a unitary UC protocol exists iff. a unitary universally-discharging (UD) protocol also exists.
To see this, we note that if $U$ is a UC protocol, then $U^\dagger$ is a UD protocol, since
\bseq \label{eq:UC_iff_UD}
\begin{align}
    \Tr [ H_B (U\rho U^\dagger - \rho)] &\geq 0 \quad \forall \rho \\
    \iff \Tr [H_B ( U (U^\dagger \rho U) U^\dagger - (U^\dagger \rho U))] & \geq 0 \quad \forall \rho \\
    \iff \Tr [H_B (\rho - U^\dagger \rho U )] &\geq 0 \quad \forall \rho \\
    \iff \Tr [H_B (V \rho V^\dagger - \rho)] & \leq 0 \quad \forall \rho, \\
    \text{with }V &= U^\dagger.\nonumber
\end{align}
\eseq
Thus, it follows that there does not exist a universally-discharging protocol if $\dim \h_B$ is finite.

We will now present a second, more illuminating proof based on Haar averages, for the non-existence of UC and UD unitary protocols for finite dimensional $\h_B$.
To that end, we show that for any arbitrary unitary $U$, the average of $\Delta E$ over all unitary rotations of any $\rho_S$ equals zero, as follows:
\bseq \label{eq:Haar_average_over_density_matrices}
\begin{align}
    \ov{\Delta E} &= \ov{\Tr [H_B (U \{ G \rho G^\dagger\} U^\dagger - G \rho G^\dagger )]} \\
    &= \frac{1}{\dim \h_B}\Tr [H_B (U \{ \Tr \rho \} \eye U^\dagger - \{ \Tr \rho\} \eye )] \\
    &= \frac{1}{\dim \h_B}\Tr [H_B (U U^\dagger - \eye)] =0.
\end{align}
\eseq
Here, the overbar denotes a Haar-average over unitary matrices $G$.
(We stress here that the above is applicable only if $\dim \h_B$ is finite.)
Thus, if $\Delta E (\rho_1) >0$ for some $\rho_1$, then there necessarily exist a $\rho_2$ for which $\Delta E (\rho_2) <0$ (and vice versa).
Since Eq.~\eqref{eq:Haar_average_over_density_matrices} is valid for all $U$ as well as all $\rho_S$, it follows that there cannot exist any UC or UD unitary protocol.
In fact, we proved a stronger statement, which says that the amount charged, averaged over all density matrices, is identically zero, for all unitary protocols~\footnote{While this equation also appeared in Ref.~\cite{Caravelli2020}, its implications were not fully appreciated therein.}. 

\begin{figure}[ht]
    \centering
    \includegraphics[width=0.65\linewidth]{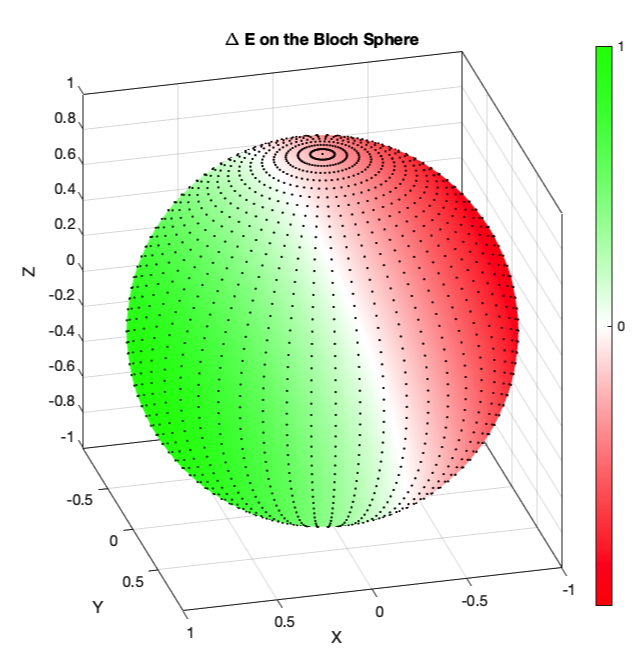}
    \caption{$\Delta E(\rho)$ from Eq.~\eqref{eq:delta_e_qubit} for a two-level system, with an arbitrary $U$, plotted over the Bloch sphere. 
    For one hemisphere, $\Delta E<0$, while for the other half, $\Delta E>0$. 
    The no-charging theorem implies that these two volumes compensate each other in general.
    Changing the unitary $U$ will only rotate the plane dividing these regions.
    }
    \label{fig:bloch}
\end{figure}
The no-charging theorem can be intuitively understood by noting that unitary protocols are essentially basis transformations.
Thus, the evolution of any initial state upon repeated identical unitary transformations cannot have `sources' or `sinks'. 
However, UC protocols effectively shift the weights of any initial state higher in energy, suggesting that repeated applications should result in a drift towards one or more high-energy `attractor' states in the space of all density matrices.
Since this is impossible for unitary transformations, the impossibility of constructing UC or UD protocols using them follows.

To illustrate Eq.~\eqref{eq:Haar_average_over_density_matrices}, we present a simple example of a 2-level quantum battery, with $H_B \ket 0 =0$ and $H_B \ket 1 = \ket 1 $.
The most general $2 \times 2$ unitary can be written as
\begin{align}
    U=\begin{pmatrix}
    a & b \\
    -b^* e^{i\phi} &a^* e^{i\phi}
    \end{pmatrix}.
\end{align}
with $|a|^2+|b|^2=1$.
% Let us look at the work given an arbitrary density matrix.
For a density matrix $\rho_0$, we have
\begin{align}
    \Delta E=\Tr[H_B(U^\dagger \rho_0 U-\rho_0)]= \bra 1 U^\dagger \rho_0 U-\rho_0 \ket 1.
\end{align}
The most general $2\times 2$ density matrix has the form
\begin{align}
    \rho_0=\begin{pmatrix}
        |\alpha|^2 & \alpha \beta^*\\
        \alpha^* \beta & |\beta|^2
    \end{pmatrix}.
\end{align}
Clearly,
\begin{align}
    \bsplit
    \Delta E&=(\left| a\right| ^2-1) \left| \beta \right| ^2 +\\
    & \left| \alpha \right| ^2 \left| b\right| ^2+2 \cos(\phi)\Re (a \alpha  b  \beta ^*). \label{eq:delta_e_qubit}
    \esplit
\end{align}
It is easy to see that if $\alpha =0$, then $\Delta E \leq 0$ for all $U$. 
(That is, the excited state can never be charged positively.)
For a two-level system, any density matrix can be described by a point on or inside the Bloch sphere.
As shown in Fig.~\ref{fig:bloch}, we find that $\Delta E >0$ and $\Delta E <0$ correspond to equal volumes.
For quantum batteries with $\dim \h_B >2$, we similarly expect that $\Delta E >0$ and $\Delta E<0$ correspond to equal volumes if $H_B$ has an equally-spaced spectrum.

\emph{Non-unitary universally-charging protocols.}---
The no-go theorem can be evaded if the protocol is non-unitary, therefore requiring interactions with the environment.
Undesirable interactions with a heat bath are usually UD.
Therefore, in theory, the existence of UC protocols seems plausible, if appropriately-designed environmental interactions are intentionally introduced.
Here, we present an explicit protocol [by specifying $H(t)$] which is UC for any value of $\dim \h_B$, by coupling it to an auxiliary qubit.
The protocol has some interesting and useful properties.
First, the Hamiltonian $H(t)$ that generates the protocol is local in the energy basis of $H_B$, a requirement for physical Hamiltonians (see also discussion below regarding double-sided energy ladders).
Second, the protocol can be tuned between the extreme limits of UC or UD by tuning the state of the control qubit.
Finally, for the UC protocol, the value of $\Delta E$ is positive for every density matrix except for the highest-charged state.

The protocol is specified by a Hamiltonian $H(t)$ that acts on $\h_\text{total}=\h_B \otimes \h_E$, with $\h_E$ denoting to the auxiliary qubit Hilbert space.
We denote the Pauli matrices by $\sigma_{x,y,z}$, while $\sigma_\pm \coloneqq (\sigma_x \pm i\sigma_y)/2$. 
The eigenvectors of $\sigma_z$ corresponding to eigenvalues $+1$ and $-1$ are denoted by $\ket \uparrow$ and $\ket \downarrow$ respectively.
Let $\dim \h_B = N$.
We consider an $H_B$ with non-degenerate eigenvalues $E_i$, with eigenvectors $\ket i$ for $i=1,\dotsc, N$, even though the model can be modified for the degenerate case.
We define a `shift operator' $s$, similar to the truncated version of the annihilation ladder operator for simple harmonic oscillators, acting on the basis states as 
\begin{align} \label{eq:energy_shift_finite_ladder}
    s\ket{n} = \begin{cases}
        \ket{n-1} & \text{if }n \in \{2,\dotsc N\} \\
        0 & \text{if } n = 1.
    \end{cases}
\end{align}
The driving protocol $H(t)$ is defined for $t\in [0,1]$, and is split into two time steps:
\begin{align}
    H(t) = \begin{cases}
        -\pi \left( \eye \otimes \sigma_x \right)&  0 \leq t < \frac{1}{2}, \\
        {\pi} \left(s^\dagger \otimes \sigma_+ + s \otimes \sigma_- \right) &  \frac{1}{2} \leq t \leq 1. \\
    \end{cases} \label{eq:protocol}
\end{align}
The corresponding time evolution operator that maps from $t=0$ to $t=1$, is given by $U = \torder \exp(-i \int_0^{1} H(t) \dd t)$, where $\torder$ denotes the time-ordering operation.

\begin{figure}
    \centering
    \includegraphics[width=0.65\linewidth]{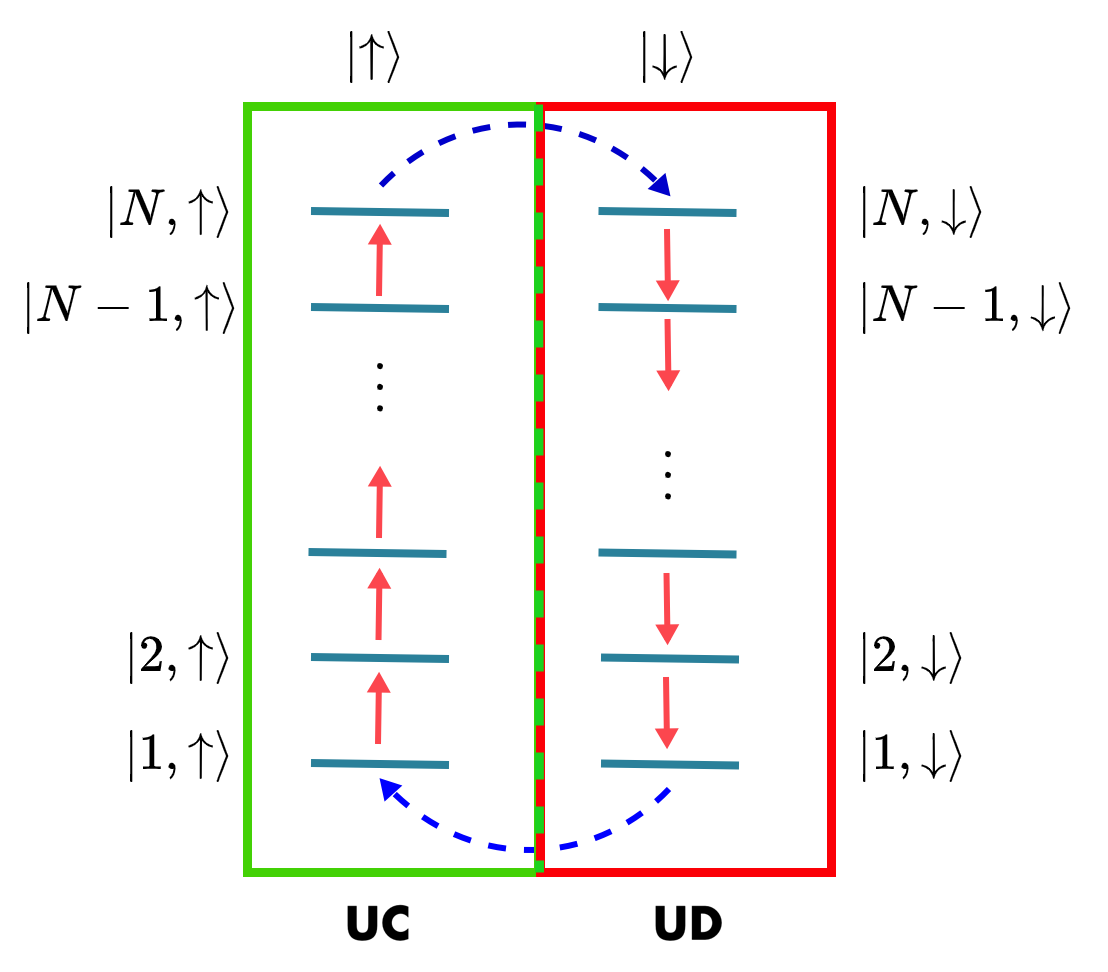}
    \caption{Pictorial representation of the unitary operator generated by the protocol~\eqref{eq:protocol}. 
    When the control qubit is set to $\ket{\uparrow}$ ($\ket{\downarrow}$), the protocol is UC (UD).}
    \label{fig:ucud}
\end{figure}

The action of $U$ on the various basis states of $\h_{\text{total}}$ is shown in Fig.~\ref{fig:ucud}. 
It is straightforward to see that when the control qubit is set to $\ket{\uparrow}$, the map for the battery states can be expressed in its Kraus form \cite{Breuer2007} as
\begin{align} 
    \map_{\ket{\uparrow}} &= s^\dagger \rho s + \Pi_{N} \rho \Pi_{N}, \label{eq:map_uparrow}
\end{align}
where $\Pi_N = \ket{N}\bra{N}$ is the projector onto the $N^{th}$ eigenstate.
(See Appendix~\ref{app:aux_qubit} for a derivation.)
$\Delta E$ corresponding to any state $\Pi_n = \ket n \bra n$ is
\begin{align}
   & \Delta E(\Pi_n) = \Tr[ H_B  (\map_{\ket{\uparrow}} [\Pi_n] - \Pi_n) ] \nonumber \\
   &= \begin{cases}
        E_{n+1} - E_n & \text{if } n \in \{ 1,\dotsc, N-1\} \\
        0 & \text{if } n =N.
    \end{cases} \geq 0.
\end{align}
Clearly, this protocol is UC, and furthermore results in a positive energy change for any initial state of the battery, except for $\rho = \Pi_N$, for which the energy is unchanged.
Furthermore, if the control qubit is set to $\ket{\downarrow}$, the protocol is UD, and the corresponding map assumes the form:
\begin{align} \label{eq:map_downarrow}
    \map_{\ket \downarrow} = s \rho s^\dagger + \Pi_1 \rho \Pi_1.
\end{align}
Interpolating the control qubit between $\ket{\uparrow}$ and $\ket \downarrow$ interpolates the protocol between these two limits.
Unlike for unitary protocols, for which, $\ov{\Delta E}=0$ as shown previously, for general CPTP maps, $\ov{\Delta E}\neq 0$. 
Indeed, for $\map _{\ket \uparrow}$, we find that $\ov{\Delta E} = E_N - E_1 > 0$, as shown in Appendix~\ref{app:Haar_CPTP}.

\emph{Batteries with infinite dimensional Hilbert spaces.}---
The two proofs for the no-charging theorem presented above are inapplicable when $\dim \h_B$ is infinite.
We now consider cases where $H_B$ has discrete and countably infinite energy levels.
Again, we assume that each energy level is non-degenerate, although our arguments can be easily extended to the cases with finite degeneracies.

Three distinct cases arise in this context: (i) $H_B$ with an energy ladder $E_0 > E_{-1} > \dotsc$, (ii) $H_B$ with an energy ladder $E_1 < E_2 < E_3 \dotsc$, and (iii), a double-sided energy ladder with energies $\dotsc E_{-1} < E_0 < E_1 \dotsc$.
We will show that for the first two cases, there do not exist any UC or UD unitary protocols, while for the third UC and UD unitary protocols do exist.
For all cases, we denote the eigenvector corresponding to $E_n$ by $\ket n$.

Let us consider the first case, in which there is a finite largest energy $E_0$.
The non-existence of a UC protocol can be proven similarly to the case of finite dimensional $\h_B$. 
Specifically, we can show that if $U$ is a UC unitary protocol, then $U\ket{k} = e^{i\phi_k} \ket k$ for all $k$, and consequently, $[U,H_B]=0$.
This implies that $\Delta E =0$ for all $\rho_B$, in contradiction with the assumption.

The second case is similar to the spectrum of a simple harmonic oscillator, but without requiring the energies to be equally spaced.
The impossibility of a UC unitary protocol for this case follows immediately from the first case.
Let us assume that $U$ is a UC protocol with respect to $H_B$ of type (ii).
From Eq.~\eqref{eq:UC_iff_UD}, we conclude that $V=U^\dagger$ must be a UD protocol, so that $\Tr [H_B (V^\dagger \rho_B V - \rho_B)] \leq 0 \ \forall \rho_B$. This implies that $\Tr [H_B' (V^\dagger \rho_B V -\rho_B)] \geq 0 \ \forall \rho_B$, where $H_B' = -H_B$ is a Hamiltonian of type (i).
Therefore, a UC unitary protocol for type (ii) exists iff. one also exists for type (i).
It therefore follows that there are no UC unitary protocols for case (ii).
The non-existence of unitary UD protocols also follows similarly for the two cases.

In contrast to all cases considered so far, we note that unitary UC protocols \textit{do exist} for case (iii), which has a double sided energy ladder.
The simplest example of a unitary UC protocol is generated by the `shift' or energy translation operator 
\begin{align}
    T = s^\dagger = \sum_{j=-\infty}^\infty \ket{j+1}\bra{j}, \label{eq:defn_energy_shift_infinite}
\end{align} 
which is simply the conjugate of the double-sided, un-truncated version of Eq.~\eqref{eq:energy_shift_finite_ladder},
In fact, $T$ corresponds to a \textit{strictly} universally-charging protocol, since $\Delta E = \Tr [H_B(s^\dagger \rho s -\rho)] >0$ for all initial density matrices.
Let us note that similar energy-shift operators have been employed in proposals of catalytic coherence~\cite{AbergCatalytic2014} where it is implemented in an auxiliary `weight' system~\cite{SkrzypczykExtracting2013,SkrzypczykWork2014,Lobejkotight2021}.

Although $T$ is a unitary UC protocol, we will now show that it cannot be `locally generated', i.e. generated by a Hamiltonian $H(t)$ that is local in the eigenbasis of $H_B$, if the battery is closed.
We recall that a local unitary $U$ is said to be locally generated~\cite{LiuClassification2023} if there exists some $H(t)$ for $0<t<1$ s.t.
\bseq \label{eq:local_generation}
\begin{align} 
    U = \torder \exp(-i\int_0^1 H(t) \dd t) \\
    \text{with } \bra{n} H(t) \ket{n'} \leq C e^{-\abs{n-n'}/l}, \label{eq:locality_condition}
\end{align}
\eseq
for all integers $n,n'$ and some constants $l,C$.
We expect the locality condition Eq.~\eqref{eq:locality_condition} to be a requirement for all ``physical'' Hamiltonians~\footnote{The double-sided energy ladder itself might be regarded as being unphysical because it lacks a ground state. However, we show that even if such a spectrum were assumed to exist, a UC protocol can still not be generated.}.

Local unitaries such as $T$ from Eq.~\eqref{eq:defn_energy_shift_infinite} can be interpreted as `one dimensional quantum walks' operating in the energy space of $H_B$.
Distinct topological classes of unitaries are distinguished by an integer-valued topological invariant $\nu$ called the flow index~\cite{KitaevAnyons2006,GrossIndex2012}, which in our context can be expressed as 
\begin{align}
    \nu (U) = \sum_{x \geq 0 > y} [U_{xy}^* U_{xy} - U_{yx}^* U_{yx}],
\end{align}
where $U_{xy}\coloneqq \bra y U \ket x$, and $\circ^*$ denotes complex conjugation. For the case of degeneracies, the former is replaced by matrices and the latter is replaced by Hermitian conjugation, and a trace operation is introduced.

Importantly, unitary operators with $\nu(U)\neq 0$ cannot be generated locally~\cite{GrossIndex2012}, i.e. no $H(t)$ satisfying Eq.~\eqref{eq:local_generation} exists.
Since $\nu(T)=1$, as can be verified easily, it follows that $T$ cannot be generated locally.
We also note that the flow index captures the net amount of upward shifting in energy of the weight of a density matrix. 
Consequently, we expect that unitary UC protocols that are more general than $T$ should also have a non-zero flow index. 
It then follows that such protocols cannot be physically realized.

However, a catalyst can help overcome this topological obstruction, as we now illustrate with an example.
To this end, we consider the auxiliary qubit model from Eq.~\eqref{eq:protocol} for the double-sided ladder, but with the shift operator defined according to Eq.~\eqref{eq:defn_energy_shift_infinite}.
It is straightforward to show that the control-qubit dependent maps [cf. Eqs.~\eqref{eq:map_uparrow}~\eqref{eq:map_downarrow}] on the battery are:
\bseq
\begin{align}
    \map_{\ket{\uparrow}} &= s^\dagger \rho s \\
    \map_{\ket{\downarrow}} &= s \rho s^\dagger,
\end{align}
\eseq
so that the qubit becomes a catalyst. Clearly, $\map_{\ket \uparrow}$ is a UC \textit{unitary} protocol corresponding to the energy-shift operator $T$.
The reason for the local generatability with the catalyst is that the unitary operator in the full Hilbert space $\h_\text{total}$, which can be pictorially represented similarly to Fig.~\ref{fig:ucud}, has a flow index of $0$, since there is no \textit{net} shift of energy in the $\h_\text{total}$.

\emph{Conclusions.}---
Fast charging protocols for quantum batteries are desirable for achieving higher efficiency.
However, charging protocols are usually designed with certain initial states in mind, and are sensitive to deviations from that state.
Such variations can be expected to arise due to undesirable environmental interactions and imperfect state preparation.
In this letter, we introduced the concept of a universally-charging (UC) protocol for quantum batteries.
These protocols are defined by their property of increasing or maintaining the average battery energy for all initial states, without ever decreasing it.
UC protocols thus exemplify charging protocols robust to initial state variations.

We proved a no-go theorem that highlights a fundamental limitation in the design of such universal protocols for quantum batteries. 
Specifically, we showed that for closed quantum batteries with finite dimensional Hilbert spaces, it is impossible to design UC protocols.
Similarly to the case of the no-cloning theorem, the no-charging theorem restricts the ability to charge (or discharge) without observation or the use of auxiliary degrees of freedom.
The implications of this theorem are profound for the development of quantum energy storage technologies, and they suggest that state observation is necessary in a fully quantum system, or the coupling with external degrees of freedom.
Observations, or partial observations, in the energy eigenbasis of the battery Hamiltonian, might in fact destroy coherence, and thus limit the effective advantage obtained by quantum correlations.

Furthermore, we showed that the amount of energy charged averaged over all initial states is identically zero, highlighting the fragility of charging options available for closed quantum batteries.
While UC protocols do exist for closed quantum batteries with double-sided infinite energy ladders, we argued that a topological obstruction makes these protocols unachievable via physical processes.
Nonetheless, the no-go theorem as well as the topological obstruction can be overcome for open quantum batteries, for which we presented an explicit recipe for constructing UC protocols.

Various research directions are suggested by our work.
It would be useful to define and understand other, more appealing forms of robustness of charging protocols.
The importance of open quantum system effects for designing UC protocols suggests that these might also be important while considering other measures of robustness.
It would be interesting to prove our conjecture that UC unitary operators have a non-zero topological index (for the case of an infinite double-sided energy ladder), since it has direct implications on the physicality or such protocols.
Understanding the trade-offs between robustness to deviations in initial states, and the speed of charging with optimized charging protocols would be a practical and important research direction.

\emph{Acknowledgments.}--
We thank Tanmoy Biswas and Luis Pedro Garc\'ia Pintos for useful discussions.
The authors acknowledge the support of NNSA for the U.S. DoE at LANL under Contract No. DE-AC52-06NA25396, and Laboratory Directed Research and Development (LDRD) for support through 20240032DR.
LANL is managed by Triad National Security, LLC, for the National Nuclear Security Administration of the U.S. DOE under contract 89233218CNA000001.

%\bibliography{biblio}

\begin{thebibliography}{51}%
\makeatletter
\providecommand \@ifxundefined [1]{%
 \@ifx{#1\undefined}
}%
\providecommand \@ifnum [1]{%
 \ifnum #1\expandafter \@firstoftwo
 \else \expandafter \@secondoftwo
 \fi
}%
\providecommand \@ifx [1]{%
 \ifx #1\expandafter \@firstoftwo
 \else \expandafter \@secondoftwo
 \fi
}%
\providecommand \natexlab [1]{#1}%
\providecommand \enquote  [1]{``#1''}%
\providecommand \bibnamefont  [1]{#1}%
\providecommand \bibfnamefont [1]{#1}%
\providecommand \citenamefont [1]{#1}%
\providecommand \href@noop [0]{\@secondoftwo}%
\providecommand \href [0]{\begingroup \@sanitize@url \@href}%
\providecommand \@href[1]{\@@startlink{#1}\@@href}%
\providecommand \@@href[1]{\endgroup#1\@@endlink}%
\providecommand \@sanitize@url [0]{\catcode `\\12\catcode `\$12\catcode
  `\&12\catcode `\#12\catcode `\^12\catcode `\_12\catcode `\%12\relax}%
\providecommand \@@startlink[1]{}%
\providecommand \@@endlink[0]{}%
\providecommand \url  [0]{\begingroup\@sanitize@url \@url }%
\providecommand \@url [1]{\endgroup\@href {#1}{\urlprefix }}%
\providecommand \urlprefix  [0]{URL }%
\providecommand \Eprint [0]{\href }%
\providecommand \doibase [0]{https://doi.org/}%
\providecommand \selectlanguage [0]{\@gobble}%
\providecommand \bibinfo  [0]{\@secondoftwo}%
\providecommand \bibfield  [0]{\@secondoftwo}%
\providecommand \translation [1]{[#1]}%
\providecommand \BibitemOpen [0]{}%
\providecommand \bibitemStop [0]{}%
\providecommand \bibitemNoStop [0]{.\EOS\space}%
\providecommand \EOS [0]{\spacefactor3000\relax}%
\providecommand \BibitemShut  [1]{\csname bibitem#1\endcsname}%
\let\auto@bib@innerbib\@empty
%</preamble>
\bibitem [{\citenamefont {Daley}\ \emph {et~al.}(2022)\citenamefont {Daley},
  \citenamefont {Bloch}, \citenamefont {Kokail}, \citenamefont {Flannigan},
  \citenamefont {Pearson}, \citenamefont {Troyer},\ and\ \citenamefont
  {Zoller}}]{DaleyPractical2022}%
  \BibitemOpen
  \bibfield  {author} {\bibinfo {author} {\bibfnamefont {A.~J.}\ \bibnamefont
  {Daley}}, \bibinfo {author} {\bibfnamefont {I.}~\bibnamefont {Bloch}},
  \bibinfo {author} {\bibfnamefont {C.}~\bibnamefont {Kokail}}, \bibinfo
  {author} {\bibfnamefont {S.}~\bibnamefont {Flannigan}}, \bibinfo {author}
  {\bibfnamefont {N.}~\bibnamefont {Pearson}}, \bibinfo {author} {\bibfnamefont
  {M.}~\bibnamefont {Troyer}},\ and\ \bibinfo {author} {\bibfnamefont
  {P.}~\bibnamefont {Zoller}},\ }\bibfield  {title} {\bibinfo {title}
  {Practical quantum advantage in quantum simulation},\ }\href
  {https://doi.org/10.1038/s41586-022-04940-6} {\bibfield  {journal} {\bibinfo
  {journal} {Nature}\ }\textbf {\bibinfo {volume} {607}},\ \bibinfo {pages}
  {667} (\bibinfo {year} {2022})}\BibitemShut {NoStop}%
\bibitem [{\citenamefont {Grumbling}\ and\ \citenamefont
  {Horowitz}(2019)}]{GrumblingQuantum2019}%
  \BibitemOpen
  \bibinfo {editor} {\bibfnamefont {E.}~\bibnamefont {Grumbling}}\ and\
  \bibinfo {editor} {\bibfnamefont {M.}~\bibnamefont {Horowitz}},\ eds.,\ \href
  {https://doi.org/10.17226/25196} {\emph {\bibinfo {title} {Quantum
  {{Computing}}: {{Progress}} and {{Prospects}}}}}\ (\bibinfo  {publisher}
  {National Academies Press},\ \bibinfo {address} {Washington, D.C.},\ \bibinfo
  {year} {2019})\BibitemShut {NoStop}%
\bibitem [{\citenamefont {Kantsepolsky}\ \emph {et~al.}(2023)\citenamefont
  {Kantsepolsky}, \citenamefont {Aviv}, \citenamefont {Weitzfeld},\ and\
  \citenamefont {Bordo}}]{KantsepolskyExploring2023}%
  \BibitemOpen
  \bibfield  {author} {\bibinfo {author} {\bibfnamefont {B.}~\bibnamefont
  {Kantsepolsky}}, \bibinfo {author} {\bibfnamefont {I.}~\bibnamefont {Aviv}},
  \bibinfo {author} {\bibfnamefont {R.}~\bibnamefont {Weitzfeld}},\ and\
  \bibinfo {author} {\bibfnamefont {E.}~\bibnamefont {Bordo}},\ }\bibfield
  {title} {\bibinfo {title} {Exploring {{Quantum Sensing Potential}} for
  {{Systems Applications}}},\ }\href
  {https://doi.org/10.1109/ACCESS.2023.3262506} {\bibfield  {journal} {\bibinfo
   {journal} {IEEE Access}\ }\textbf {\bibinfo {volume} {11}},\ \bibinfo
  {pages} {31569} (\bibinfo {year} {2023})}\BibitemShut {NoStop}%
\bibitem [{\citenamefont {Degen}\ \emph {et~al.}(2017)\citenamefont {Degen},
  \citenamefont {Reinhard},\ and\ \citenamefont
  {Cappellaro}}]{DegenQuantum2017}%
  \BibitemOpen
  \bibfield  {author} {\bibinfo {author} {\bibfnamefont {C.~L.}\ \bibnamefont
  {Degen}}, \bibinfo {author} {\bibfnamefont {F.}~\bibnamefont {Reinhard}},\
  and\ \bibinfo {author} {\bibfnamefont {P.}~\bibnamefont {Cappellaro}},\
  }\bibfield  {title} {\bibinfo {title} {Quantum sensing},\ }\href
  {https://doi.org/10.1103/RevModPhys.89.035002} {\bibfield  {journal}
  {\bibinfo  {journal} {Reviews of Modern Physics}\ }\textbf {\bibinfo {volume}
  {89}},\ \bibinfo {pages} {035002} (\bibinfo {year} {2017})}\BibitemShut
  {NoStop}%
\bibitem [{\citenamefont {Renner}\ and\ \citenamefont
  {Wolf}(2023)}]{RennerQuantum2023}%
  \BibitemOpen
  \bibfield  {author} {\bibinfo {author} {\bibfnamefont {R.}~\bibnamefont
  {Renner}}\ and\ \bibinfo {author} {\bibfnamefont {R.}~\bibnamefont {Wolf}},\
  }\bibfield  {title} {\bibinfo {title} {Quantum {{Advantage}} in
  {{Cryptography}}},\ }\href {https://doi.org/10.2514/1.J062267} {\bibfield
  {journal} {\bibinfo  {journal} {AIAA Journal}\ }\textbf {\bibinfo {volume}
  {61}},\ \bibinfo {pages} {1895} (\bibinfo {year} {2023})}\BibitemShut
  {NoStop}%
\bibitem [{\citenamefont {Alicki}\ and\ \citenamefont
  {Fannes}(2013)}]{AlickiEntanglement2013}%
  \BibitemOpen
  \bibfield  {author} {\bibinfo {author} {\bibfnamefont {R.}~\bibnamefont
  {Alicki}}\ and\ \bibinfo {author} {\bibfnamefont {M.}~\bibnamefont
  {Fannes}},\ }\bibfield  {title} {\bibinfo {title} {Entanglement boost for
  extractable work from ensembles of quantum batteries},\ }\href
  {https://doi.org/10.1103/PhysRevE.87.042123} {\bibfield  {journal} {\bibinfo
  {journal} {Physical Review E}\ }\textbf {\bibinfo {volume} {87}},\ \bibinfo
  {pages} {042123} (\bibinfo {year} {2013})},\ \bibinfo {note}
  {\url{https://link.aps.org/doi/10.1103/PhysRevE.87.042123}}\BibitemShut
  {NoStop}%
\bibitem [{\citenamefont {Bhattacharjee}\ and\ \citenamefont
  {Dutta}(2021)}]{BhattacharjeeQuantum2021}%
  \BibitemOpen
  \bibfield  {author} {\bibinfo {author} {\bibfnamefont {S.}~\bibnamefont
  {Bhattacharjee}}\ and\ \bibinfo {author} {\bibfnamefont {A.}~\bibnamefont
  {Dutta}},\ }\bibfield  {title} {\bibinfo {title} {Quantum thermal machines
  and batteries},\ }\href {https://doi.org/10.1140/epjb/s10051-021-00235-3}
  {\bibfield  {journal} {\bibinfo  {journal} {The European Physical Journal B}\
  }\textbf {\bibinfo {volume} {94}},\ \bibinfo {pages} {239} (\bibinfo {year}
  {2021})},\ \bibinfo {note}
  {\url{https://doi.org/10.1140/epjb/s10051-021-00235-3}}\BibitemShut {NoStop}%
\bibitem [{\citenamefont {Auffèves}(2022)}]{Auffves2022}%
  \BibitemOpen
  \bibfield  {author} {\bibinfo {author} {\bibfnamefont {A.}~\bibnamefont
  {Auffèves}},\ }\bibfield  {title} {\bibinfo {title} {Quantum technologies
  need a quantum energy initiative},\ }\bibfield  {journal} {\bibinfo
  {journal} {PRX Quantum}\ }\textbf {\bibinfo {volume} {3}},\ \href
  {https://doi.org/10.1103/prxquantum.3.020101} {10.1103/prxquantum.3.020101}
  (\bibinfo {year} {2022})\BibitemShut {NoStop}%
\bibitem [{\citenamefont {Campaioli}\ \emph {et~al.}(2018)\citenamefont
  {Campaioli}, \citenamefont {Pollock},\ and\ \citenamefont
  {Vinjanampathy}}]{campaioli}%
  \BibitemOpen
  \bibfield  {author} {\bibinfo {author} {\bibfnamefont {F.}~\bibnamefont
  {Campaioli}}, \bibinfo {author} {\bibfnamefont {F.~A.}\ \bibnamefont
  {Pollock}},\ and\ \bibinfo {author} {\bibfnamefont {S.}~\bibnamefont
  {Vinjanampathy}},\ }\href {https://doi.org/10.1007/978-3-319-99046-0} {\emph
  {\bibinfo {title} {Quantum Batteries. In: Thermodynamics in the Quantum
  Regime: Fundamental Aspects and New Directions}}}\ (\bibinfo  {publisher}
  {Springer, Cham},\ \bibinfo {year} {2018})\BibitemShut {NoStop}%
\bibitem [{\citenamefont {Ferraro}\ \emph
  {et~al.}(2018{\natexlab{a}})\citenamefont {Ferraro}, \citenamefont {Campisi},
  \citenamefont {Andolina}, \citenamefont {Pellegrini},\ and\ \citenamefont
  {Polini}}]{FerraroHighPower2018}%
  \BibitemOpen
  \bibfield  {author} {\bibinfo {author} {\bibfnamefont {D.}~\bibnamefont
  {Ferraro}}, \bibinfo {author} {\bibfnamefont {M.}~\bibnamefont {Campisi}},
  \bibinfo {author} {\bibfnamefont {G.~M.}\ \bibnamefont {Andolina}}, \bibinfo
  {author} {\bibfnamefont {V.}~\bibnamefont {Pellegrini}},\ and\ \bibinfo
  {author} {\bibfnamefont {M.}~\bibnamefont {Polini}},\ }\bibfield  {title}
  {\bibinfo {title} {High-{{Power Collective Charging}} of a {{Solid-State
  Quantum Battery}}},\ }\href {https://doi.org/10.1103/PhysRevLett.120.117702}
  {\bibfield  {journal} {\bibinfo  {journal} {Physical Review Letters}\
  }\textbf {\bibinfo {volume} {120}},\ \bibinfo {pages} {117702} (\bibinfo
  {year} {2018}{\natexlab{a}})},\ \bibinfo {note}
  {\url{https://link.aps.org/doi/10.1103/PhysRevLett.120.117702}}\BibitemShut
  {NoStop}%
\bibitem [{\citenamefont {Salvia}\ \emph {et~al.}(2023)\citenamefont {Salvia},
  \citenamefont {Perarnau-Llobet}, \citenamefont {Haack}, \citenamefont
  {Brunner},\ and\ \citenamefont {Nimmrichter}}]{Salvia2023}%
  \BibitemOpen
  \bibfield  {author} {\bibinfo {author} {\bibfnamefont {R.}~\bibnamefont
  {Salvia}}, \bibinfo {author} {\bibfnamefont {M.}~\bibnamefont
  {Perarnau-Llobet}}, \bibinfo {author} {\bibfnamefont {G.}~\bibnamefont
  {Haack}}, \bibinfo {author} {\bibfnamefont {N.}~\bibnamefont {Brunner}},\
  and\ \bibinfo {author} {\bibfnamefont {S.}~\bibnamefont {Nimmrichter}},\
  }\bibfield  {title} {\bibinfo {title} {Quantum advantage in charging cavity
  and spin batteries by repeated interactions},\ }\bibfield  {journal}
  {\bibinfo  {journal} {Physical Review Research}\ }\textbf {\bibinfo {volume}
  {5}},\ \href {https://doi.org/10.1103/physrevresearch.5.013155}
  {10.1103/physrevresearch.5.013155} (\bibinfo {year} {2023})\BibitemShut
  {NoStop}%
\bibitem [{\citenamefont {Seah}\ \emph {et~al.}(2021)\citenamefont {Seah},
  \citenamefont {Perarnau-Llobet}, \citenamefont {Haack}, \citenamefont
  {Brunner},\ and\ \citenamefont {Nimmrichter}}]{Seah2021}%
  \BibitemOpen
  \bibfield  {author} {\bibinfo {author} {\bibfnamefont {S.}~\bibnamefont
  {Seah}}, \bibinfo {author} {\bibfnamefont {M.}~\bibnamefont
  {Perarnau-Llobet}}, \bibinfo {author} {\bibfnamefont {G.}~\bibnamefont
  {Haack}}, \bibinfo {author} {\bibfnamefont {N.}~\bibnamefont {Brunner}},\
  and\ \bibinfo {author} {\bibfnamefont {S.}~\bibnamefont {Nimmrichter}},\
  }\bibfield  {title} {\bibinfo {title} {Quantum speed-up in collisional
  battery charging},\ }\bibfield  {journal} {\bibinfo  {journal} {Physical
  Review Letters}\ }\textbf {\bibinfo {volume} {127}},\ \href
  {https://doi.org/10.1103/physrevlett.127.100601}
  {10.1103/physrevlett.127.100601} (\bibinfo {year} {2021})\BibitemShut
  {NoStop}%
\bibitem [{\citenamefont {Andolina}\ \emph {et~al.}(2018)\citenamefont
  {Andolina}, \citenamefont {Farina}, \citenamefont {Mari}, \citenamefont
  {Pellegrini}, \citenamefont {Giovannetti},\ and\ \citenamefont
  {Polini}}]{PoliniPRB2018}%
  \BibitemOpen
  \bibfield  {author} {\bibinfo {author} {\bibfnamefont {G.~M.}\ \bibnamefont
  {Andolina}}, \bibinfo {author} {\bibfnamefont {D.}~\bibnamefont {Farina}},
  \bibinfo {author} {\bibfnamefont {A.}~\bibnamefont {Mari}}, \bibinfo {author}
  {\bibfnamefont {V.}~\bibnamefont {Pellegrini}}, \bibinfo {author}
  {\bibfnamefont {V.}~\bibnamefont {Giovannetti}},\ and\ \bibinfo {author}
  {\bibfnamefont {M.}~\bibnamefont {Polini}},\ }\bibfield  {title} {\bibinfo
  {title} {Charger-mediated energy transfer in exactly solvable models for
  quantum batteries},\ }\href {https://doi.org/10.1103/PhysRevB.98.205423}
  {\bibfield  {journal} {\bibinfo  {journal} {Phys. Rev. B Condens. Matter}\
  }\textbf {\bibinfo {volume} {98}},\ \bibinfo {pages} {205423} (\bibinfo
  {year} {2018})}\BibitemShut {NoStop}%
\bibitem [{\citenamefont {Andolina}\ \emph
  {et~al.}(2019{\natexlab{a}})\citenamefont {Andolina}, \citenamefont {Keck},
  \citenamefont {Mari}, \citenamefont {Campisi}, \citenamefont {Giovannetti},\
  and\ \citenamefont {Polini}}]{PoliniPRL2019}%
  \BibitemOpen
  \bibfield  {author} {\bibinfo {author} {\bibfnamefont {G.~M.}\ \bibnamefont
  {Andolina}}, \bibinfo {author} {\bibfnamefont {M.}~\bibnamefont {Keck}},
  \bibinfo {author} {\bibfnamefont {A.}~\bibnamefont {Mari}}, \bibinfo {author}
  {\bibfnamefont {M.}~\bibnamefont {Campisi}}, \bibinfo {author} {\bibfnamefont
  {V.}~\bibnamefont {Giovannetti}},\ and\ \bibinfo {author} {\bibfnamefont
  {M.}~\bibnamefont {Polini}},\ }\bibfield  {title} {\bibinfo {title}
  {Extractable work, the role of correlations, and asymptotic freedom in
  quantum batteries},\ }\href {https://doi.org/10.1103/PhysRevLett.122.047702}
  {\bibfield  {journal} {\bibinfo  {journal} {Phys. Rev. Lett.}\ }\textbf
  {\bibinfo {volume} {122}},\ \bibinfo {pages} {047702} (\bibinfo {year}
  {2019}{\natexlab{a}})}\BibitemShut {NoStop}%
\bibitem [{\citenamefont {Andolina}\ \emph
  {et~al.}(2019{\natexlab{b}})\citenamefont {Andolina}, \citenamefont {Keck},
  \citenamefont {Mari}, \citenamefont {Giovannetti},\ and\ \citenamefont
  {Polini}}]{PoliniPRB2019}%
  \BibitemOpen
  \bibfield  {author} {\bibinfo {author} {\bibfnamefont {G.~M.}\ \bibnamefont
  {Andolina}}, \bibinfo {author} {\bibfnamefont {M.}~\bibnamefont {Keck}},
  \bibinfo {author} {\bibfnamefont {A.}~\bibnamefont {Mari}}, \bibinfo {author}
  {\bibfnamefont {V.}~\bibnamefont {Giovannetti}},\ and\ \bibinfo {author}
  {\bibfnamefont {M.}~\bibnamefont {Polini}},\ }\bibfield  {title} {\bibinfo
  {title} {Quantum versus classical many-body batteries},\ }\href
  {https://doi.org/10.1103/PhysRevB.99.205437} {\bibfield  {journal} {\bibinfo
  {journal} {Phys. Rev. B Condens. Matter}\ }\textbf {\bibinfo {volume} {99}},\
  \bibinfo {pages} {205437} (\bibinfo {year} {2019}{\natexlab{b}})}\BibitemShut
  {NoStop}%
\bibitem [{\citenamefont {Ferraro}\ \emph
  {et~al.}(2018{\natexlab{b}})\citenamefont {Ferraro}, \citenamefont {Campisi},
  \citenamefont {Andolina}, \citenamefont {Pellegrini},\ and\ \citenamefont
  {Polini}}]{Ferraro2018}%
  \BibitemOpen
  \bibfield  {author} {\bibinfo {author} {\bibfnamefont {D.}~\bibnamefont
  {Ferraro}}, \bibinfo {author} {\bibfnamefont {M.}~\bibnamefont {Campisi}},
  \bibinfo {author} {\bibfnamefont {G.~M.}\ \bibnamefont {Andolina}}, \bibinfo
  {author} {\bibfnamefont {V.}~\bibnamefont {Pellegrini}},\ and\ \bibinfo
  {author} {\bibfnamefont {M.}~\bibnamefont {Polini}},\ }\bibfield  {title}
  {\bibinfo {title} {High-power collective charging of a solid-state quantum
  battery},\ }\bibfield  {journal} {\bibinfo  {journal} {Physical Review
  Letters}\ }\textbf {\bibinfo {volume} {120}},\ \href
  {https://doi.org/10.1103/physrevlett.120.117702}
  {10.1103/physrevlett.120.117702} (\bibinfo {year}
  {2018}{\natexlab{b}})\BibitemShut {NoStop}%
\bibitem [{\citenamefont {Le}\ \emph {et~al.}(2018)\citenamefont {Le},
  \citenamefont {Levinsen}, \citenamefont {Modi}, \citenamefont {Parish},\ and\
  \citenamefont {Pollock}}]{Le2018}%
  \BibitemOpen
  \bibfield  {author} {\bibinfo {author} {\bibfnamefont {T.~P.}\ \bibnamefont
  {Le}}, \bibinfo {author} {\bibfnamefont {J.}~\bibnamefont {Levinsen}},
  \bibinfo {author} {\bibfnamefont {K.}~\bibnamefont {Modi}}, \bibinfo {author}
  {\bibfnamefont {M.~M.}\ \bibnamefont {Parish}},\ and\ \bibinfo {author}
  {\bibfnamefont {F.~A.}\ \bibnamefont {Pollock}},\ }\bibfield  {title}
  {\bibinfo {title} {Spin-chain model of a many-body quantum battery},\
  }\bibfield  {journal} {\bibinfo  {journal} {Physical Review A}\ }\textbf
  {\bibinfo {volume} {97}},\ \href {https://doi.org/10.1103/physreva.97.022106}
  {10.1103/physreva.97.022106} (\bibinfo {year} {2018})\BibitemShut {NoStop}%
\bibitem [{\citenamefont {Rossini}\ \emph
  {et~al.}(2020{\natexlab{a}})\citenamefont {Rossini}, \citenamefont
  {Andolina}, \citenamefont {Rosa}, \citenamefont {Carrega},\ and\
  \citenamefont {Polini}}]{Rossini2020}%
  \BibitemOpen
  \bibfield  {author} {\bibinfo {author} {\bibfnamefont {D.}~\bibnamefont
  {Rossini}}, \bibinfo {author} {\bibfnamefont {G.~M.}\ \bibnamefont
  {Andolina}}, \bibinfo {author} {\bibfnamefont {D.}~\bibnamefont {Rosa}},
  \bibinfo {author} {\bibfnamefont {M.}~\bibnamefont {Carrega}},\ and\ \bibinfo
  {author} {\bibfnamefont {M.}~\bibnamefont {Polini}},\ }\bibfield  {title}
  {\bibinfo {title} {Quantum advantage in the charging process of
  sachdev-ye-kitaev batteries},\ }\bibfield  {journal} {\bibinfo  {journal}
  {Physical Review Letters}\ }\textbf {\bibinfo {volume} {125}},\ \href
  {https://doi.org/10.1103/physrevlett.125.236402}
  {10.1103/physrevlett.125.236402} (\bibinfo {year}
  {2020}{\natexlab{a}})\BibitemShut {NoStop}%
\bibitem [{\citenamefont {Shaghaghi}\ \emph {et~al.}(2022)\citenamefont
  {Shaghaghi}, \citenamefont {Singh}, \citenamefont {Benenti},\ and\
  \citenamefont {Rosa}}]{Shaghaghi2022}%
  \BibitemOpen
  \bibfield  {author} {\bibinfo {author} {\bibfnamefont {V.}~\bibnamefont
  {Shaghaghi}}, \bibinfo {author} {\bibfnamefont {V.}~\bibnamefont {Singh}},
  \bibinfo {author} {\bibfnamefont {G.}~\bibnamefont {Benenti}},\ and\ \bibinfo
  {author} {\bibfnamefont {D.}~\bibnamefont {Rosa}},\ }\bibfield  {title}
  {\bibinfo {title} {Micromasers as quantum batteries},\ }\href
  {https://doi.org/10.1088/2058-9565/ac8829} {\bibfield  {journal} {\bibinfo
  {journal} {Quantum Science and Technology}\ }\textbf {\bibinfo {volume}
  {7}},\ \bibinfo {pages} {04LT01} (\bibinfo {year} {2022})}\BibitemShut
  {NoStop}%
\bibitem [{\citenamefont {Lu}\ \emph {et~al.}(2024)\citenamefont {Lu},
  \citenamefont {Tian}, \citenamefont {L\"{u}},\ and\ \citenamefont
  {Shang}}]{topquantbat}%
  \BibitemOpen
  \bibfield  {author} {\bibinfo {author} {\bibfnamefont {Z.-G.}\ \bibnamefont
  {Lu}}, \bibinfo {author} {\bibfnamefont {G.}~\bibnamefont {Tian}}, \bibinfo
  {author} {\bibfnamefont {X.-Y.}\ \bibnamefont {L\"{u}}},\ and\ \bibinfo
  {author} {\bibfnamefont {C.}~\bibnamefont {Shang}},\ }\href
  {https://doi.org/10.48550/ARXIV.2405.03675} {\bibinfo {title} {Topological
  quantum batteries}} (\bibinfo {year} {2024})\BibitemShut {NoStop}%
\bibitem [{\citenamefont {Rossini}\ \emph
  {et~al.}(2020{\natexlab{b}})\citenamefont {Rossini}, \citenamefont
  {Andolina}, \citenamefont {Rosa}, \citenamefont {Carrega},\ and\
  \citenamefont {Polini}}]{andolinasyk}%
  \BibitemOpen
  \bibfield  {author} {\bibinfo {author} {\bibfnamefont {D.}~\bibnamefont
  {Rossini}}, \bibinfo {author} {\bibfnamefont {G.~M.}\ \bibnamefont
  {Andolina}}, \bibinfo {author} {\bibfnamefont {D.}~\bibnamefont {Rosa}},
  \bibinfo {author} {\bibfnamefont {M.}~\bibnamefont {Carrega}},\ and\ \bibinfo
  {author} {\bibfnamefont {M.}~\bibnamefont {Polini}},\ }\bibfield  {title}
  {\bibinfo {title} {Quantum advantage in the charging process of
  {Sachdev-Ye-Kitaev} batteries},\ }\href
  {https://doi.org/10.1103/PhysRevLett.125.236402} {\bibfield  {journal}
  {\bibinfo  {journal} {Phys. Rev. Lett.}\ }\textbf {\bibinfo {volume} {125}},\
  \bibinfo {pages} {236402} (\bibinfo {year} {2020}{\natexlab{b}})}\BibitemShut
  {NoStop}%
\bibitem [{\citenamefont {Binder}\ \emph {et~al.}(2015)\citenamefont {Binder},
  \citenamefont {Vinjanampathy}, \citenamefont {Modi},\ and\ \citenamefont
  {Goold}}]{Binder2015}%
  \BibitemOpen
  \bibfield  {author} {\bibinfo {author} {\bibfnamefont {F.~C.}\ \bibnamefont
  {Binder}}, \bibinfo {author} {\bibfnamefont {S.}~\bibnamefont
  {Vinjanampathy}}, \bibinfo {author} {\bibfnamefont {K.}~\bibnamefont
  {Modi}},\ and\ \bibinfo {author} {\bibfnamefont {J.}~\bibnamefont {Goold}},\
  }\bibfield  {title} {\bibinfo {title} {Quantacell: powerful charging of
  quantum batteries},\ }\href {https://doi.org/10.1088/1367-2630/17/7/075015}
  {\bibfield  {journal} {\bibinfo  {journal} {New Journal of Physics}\ }\textbf
  {\bibinfo {volume} {17}},\ \bibinfo {pages} {075015} (\bibinfo {year}
  {2015})}\BibitemShut {NoStop}%
\bibitem [{\citenamefont {Juli{\`a}-Farr{\'e}}\ \emph
  {et~al.}(2020)\citenamefont {Juli{\`a}-Farr{\'e}}, \citenamefont {Salamon},
  \citenamefont {Riera}, \citenamefont {Bera},\ and\ \citenamefont
  {Lewenstein}}]{LewensteinBatteries18}%
  \BibitemOpen
  \bibfield  {author} {\bibinfo {author} {\bibfnamefont {S.}~\bibnamefont
  {Juli{\`a}-Farr{\'e}}}, \bibinfo {author} {\bibfnamefont {T.}~\bibnamefont
  {Salamon}}, \bibinfo {author} {\bibfnamefont {A.}~\bibnamefont {Riera}},
  \bibinfo {author} {\bibfnamefont {M.~N.}\ \bibnamefont {Bera}},\ and\
  \bibinfo {author} {\bibfnamefont {M.}~\bibnamefont {Lewenstein}},\ }\bibfield
   {title} {\bibinfo {title} {Bounds on the capacity and power of quantum
  batteries},\ }\href {https://doi.org/10.1103/PhysRevResearch.2.023113}
  {\bibfield  {journal} {\bibinfo  {journal} {Phys. Rev. Research}\ }\textbf
  {\bibinfo {volume} {2}},\ \bibinfo {pages} {023113} (\bibinfo {year}
  {2020})}\BibitemShut {NoStop}%
\bibitem [{\citenamefont {Bakhshinezhad}\ \emph {et~al.}(2024)\citenamefont
  {Bakhshinezhad}, \citenamefont {Jablonski}, \citenamefont {Binder},\ and\
  \citenamefont {Friis}}]{Bakhshinezhad2024}%
  \BibitemOpen
  \bibfield  {author} {\bibinfo {author} {\bibfnamefont {P.}~\bibnamefont
  {Bakhshinezhad}}, \bibinfo {author} {\bibfnamefont {B.~R.}\ \bibnamefont
  {Jablonski}}, \bibinfo {author} {\bibfnamefont {F.~C.}\ \bibnamefont
  {Binder}},\ and\ \bibinfo {author} {\bibfnamefont {N.}~\bibnamefont
  {Friis}},\ }\bibfield  {title} {\bibinfo {title} {Trade-offs between
  precision and fluctuations in charging finite-dimensional quantum
  batteries},\ }\bibfield  {journal} {\bibinfo  {journal} {Physical Review E}\
  }\textbf {\bibinfo {volume} {109}},\ \href
  {https://doi.org/10.1103/physreve.109.014131} {10.1103/physreve.109.014131}
  (\bibinfo {year} {2024})\BibitemShut {NoStop}%
\bibitem [{\citenamefont {Caravelli}\ \emph {et~al.}(2021)\citenamefont
  {Caravelli}, \citenamefont {Yan}, \citenamefont {García-Pintos},\ and\
  \citenamefont {Hamma}}]{Caravelli2021}%
  \BibitemOpen
  \bibfield  {author} {\bibinfo {author} {\bibfnamefont {F.}~\bibnamefont
  {Caravelli}}, \bibinfo {author} {\bibfnamefont {B.}~\bibnamefont {Yan}},
  \bibinfo {author} {\bibfnamefont {L.~P.}\ \bibnamefont {García-Pintos}},\
  and\ \bibinfo {author} {\bibfnamefont {A.}~\bibnamefont {Hamma}},\ }\bibfield
   {title} {\bibinfo {title} {Energy storage and coherence in closed and open
  quantum batteries},\ }\href {https://doi.org/10.22331/q-2021-07-15-505}
  {\bibfield  {journal} {\bibinfo  {journal} {Quantum}\ }\textbf {\bibinfo
  {volume} {5}},\ \bibinfo {pages} {505} (\bibinfo {year} {2021})}\BibitemShut
  {NoStop}%
\bibitem [{\citenamefont {Shi}\ \emph {et~al.}(2022)\citenamefont {Shi},
  \citenamefont {Ding}, \citenamefont {Wan}, \citenamefont {Wang},\ and\
  \citenamefont {Yang}}]{Shi2022}%
  \BibitemOpen
  \bibfield  {author} {\bibinfo {author} {\bibfnamefont {H.-L.}\ \bibnamefont
  {Shi}}, \bibinfo {author} {\bibfnamefont {S.}~\bibnamefont {Ding}}, \bibinfo
  {author} {\bibfnamefont {Q.-K.}\ \bibnamefont {Wan}}, \bibinfo {author}
  {\bibfnamefont {X.-H.}\ \bibnamefont {Wang}},\ and\ \bibinfo {author}
  {\bibfnamefont {W.-L.}\ \bibnamefont {Yang}},\ }\bibfield  {title} {\bibinfo
  {title} {Entanglement, coherence, and extractable work in quantum
  batteries},\ }\bibfield  {journal} {\bibinfo  {journal} {Physical Review
  Letters}\ }\textbf {\bibinfo {volume} {129}},\ \href
  {https://doi.org/10.1103/physrevlett.129.130602}
  {10.1103/physrevlett.129.130602} (\bibinfo {year} {2022})\BibitemShut
  {NoStop}%
\bibitem [{\citenamefont {Gyhm}\ \emph {et~al.}(2022)\citenamefont {Gyhm},
  \citenamefont {Šafránek},\ and\ \citenamefont {Rosa}}]{Gyhm2022}%
  \BibitemOpen
  \bibfield  {author} {\bibinfo {author} {\bibfnamefont {J.-Y.}\ \bibnamefont
  {Gyhm}}, \bibinfo {author} {\bibfnamefont {D.}~\bibnamefont {Šafránek}},\
  and\ \bibinfo {author} {\bibfnamefont {D.}~\bibnamefont {Rosa}},\ }\bibfield
  {title} {\bibinfo {title} {Quantum charging advantage cannot be extensive
  without global operations},\ }\bibfield  {journal} {\bibinfo  {journal}
  {Physical Review Letters}\ }\textbf {\bibinfo {volume} {128}},\ \href
  {https://doi.org/10.1103/physrevlett.128.140501}
  {10.1103/physrevlett.128.140501} (\bibinfo {year} {2022})\BibitemShut
  {NoStop}%
\bibitem [{\citenamefont {Quach}\ \emph {et~al.}(2022)\citenamefont {Quach},
  \citenamefont {McGhee}, \citenamefont {Ganzer}, \citenamefont {Rouse},
  \citenamefont {Lovett}, \citenamefont {Gauger}, \citenamefont {Keeling},
  \citenamefont {Cerullo}, \citenamefont {Lidzey},\ and\ \citenamefont
  {Virgili}}]{Quach2022}%
  \BibitemOpen
  \bibfield  {author} {\bibinfo {author} {\bibfnamefont {J.~Q.}\ \bibnamefont
  {Quach}}, \bibinfo {author} {\bibfnamefont {K.~E.}\ \bibnamefont {McGhee}},
  \bibinfo {author} {\bibfnamefont {L.}~\bibnamefont {Ganzer}}, \bibinfo
  {author} {\bibfnamefont {D.~M.}\ \bibnamefont {Rouse}}, \bibinfo {author}
  {\bibfnamefont {B.~W.}\ \bibnamefont {Lovett}}, \bibinfo {author}
  {\bibfnamefont {E.~M.}\ \bibnamefont {Gauger}}, \bibinfo {author}
  {\bibfnamefont {J.}~\bibnamefont {Keeling}}, \bibinfo {author} {\bibfnamefont
  {G.}~\bibnamefont {Cerullo}}, \bibinfo {author} {\bibfnamefont {D.~G.}\
  \bibnamefont {Lidzey}},\ and\ \bibinfo {author} {\bibfnamefont
  {T.}~\bibnamefont {Virgili}},\ }\bibfield  {title} {\bibinfo {title}
  {Superabsorption in an organic microcavity: Toward a quantum battery},\
  }\bibfield  {journal} {\bibinfo  {journal} {Science Advances}\ }\textbf
  {\bibinfo {volume} {8}},\ \href {https://doi.org/10.1126/sciadv.abk3160}
  {10.1126/sciadv.abk3160} (\bibinfo {year} {2022})\BibitemShut {NoStop}%
\bibitem [{\citenamefont {Yu}\ \emph {et~al.}(2024)\citenamefont {Yu},
  \citenamefont {Wang}, \citenamefont {Liu}, \citenamefont {Zha}, \citenamefont
  {Wu}, \citenamefont {Chen}, \citenamefont {Ye}, \citenamefont {Li},
  \citenamefont {Zhu}, \citenamefont {Guo}, \citenamefont {Qian}, \citenamefont
  {Huang}, \citenamefont {Zhao}, \citenamefont {Ying}, \citenamefont {Fan},
  \citenamefont {Wu}, \citenamefont {Su}, \citenamefont {Deng}, \citenamefont
  {Rong}, \citenamefont {Zhang}, \citenamefont {Cao}, \citenamefont {Lin},
  \citenamefont {Xu}, \citenamefont {Guo}, \citenamefont {Li}, \citenamefont
  {Liang}, \citenamefont {Wu}, \citenamefont {Huo}, \citenamefont {Lu},
  \citenamefont {Peng}, \citenamefont {Nemoto}, \citenamefont {Munro},
  \citenamefont {Zhu}, \citenamefont {Pan},\ and\ \citenamefont
  {Gong}}]{Yu2024}%
  \BibitemOpen
  \bibfield  {author} {\bibinfo {author} {\bibfnamefont {J.}~\bibnamefont
  {Yu}}, \bibinfo {author} {\bibfnamefont {S.}~\bibnamefont {Wang}}, \bibinfo
  {author} {\bibfnamefont {K.}~\bibnamefont {Liu}}, \bibinfo {author}
  {\bibfnamefont {C.}~\bibnamefont {Zha}}, \bibinfo {author} {\bibfnamefont
  {Y.}~\bibnamefont {Wu}}, \bibinfo {author} {\bibfnamefont {F.}~\bibnamefont
  {Chen}}, \bibinfo {author} {\bibfnamefont {Y.}~\bibnamefont {Ye}}, \bibinfo
  {author} {\bibfnamefont {S.}~\bibnamefont {Li}}, \bibinfo {author}
  {\bibfnamefont {Q.}~\bibnamefont {Zhu}}, \bibinfo {author} {\bibfnamefont
  {S.}~\bibnamefont {Guo}}, \bibinfo {author} {\bibfnamefont {H.}~\bibnamefont
  {Qian}}, \bibinfo {author} {\bibfnamefont {H.-L.}\ \bibnamefont {Huang}},
  \bibinfo {author} {\bibfnamefont {Y.}~\bibnamefont {Zhao}}, \bibinfo {author}
  {\bibfnamefont {C.}~\bibnamefont {Ying}}, \bibinfo {author} {\bibfnamefont
  {D.}~\bibnamefont {Fan}}, \bibinfo {author} {\bibfnamefont {D.}~\bibnamefont
  {Wu}}, \bibinfo {author} {\bibfnamefont {H.}~\bibnamefont {Su}}, \bibinfo
  {author} {\bibfnamefont {H.}~\bibnamefont {Deng}}, \bibinfo {author}
  {\bibfnamefont {H.}~\bibnamefont {Rong}}, \bibinfo {author} {\bibfnamefont
  {K.}~\bibnamefont {Zhang}}, \bibinfo {author} {\bibfnamefont
  {S.}~\bibnamefont {Cao}}, \bibinfo {author} {\bibfnamefont {J.}~\bibnamefont
  {Lin}}, \bibinfo {author} {\bibfnamefont {Y.}~\bibnamefont {Xu}}, \bibinfo
  {author} {\bibfnamefont {C.}~\bibnamefont {Guo}}, \bibinfo {author}
  {\bibfnamefont {N.}~\bibnamefont {Li}}, \bibinfo {author} {\bibfnamefont
  {F.}~\bibnamefont {Liang}}, \bibinfo {author} {\bibfnamefont
  {G.}~\bibnamefont {Wu}}, \bibinfo {author} {\bibfnamefont {Y.-H.}\
  \bibnamefont {Huo}}, \bibinfo {author} {\bibfnamefont {C.-Y.}\ \bibnamefont
  {Lu}}, \bibinfo {author} {\bibfnamefont {C.-Z.}\ \bibnamefont {Peng}},
  \bibinfo {author} {\bibfnamefont {K.}~\bibnamefont {Nemoto}}, \bibinfo
  {author} {\bibfnamefont {W.~J.}\ \bibnamefont {Munro}}, \bibinfo {author}
  {\bibfnamefont {X.}~\bibnamefont {Zhu}}, \bibinfo {author} {\bibfnamefont
  {J.-W.}\ \bibnamefont {Pan}},\ and\ \bibinfo {author} {\bibfnamefont
  {M.}~\bibnamefont {Gong}},\ }\bibfield  {title} {\bibinfo {title}
  {Experimental demonstration of a maxwell’s demon quantum battery in a
  superconducting noisy intermediate-scale quantum processor},\ }\bibfield
  {journal} {\bibinfo  {journal} {Physical Review A}\ }\textbf {\bibinfo
  {volume} {109}},\ \href {https://doi.org/10.1103/physreva.109.062614}
  {10.1103/physreva.109.062614} (\bibinfo {year} {2024})\BibitemShut {NoStop}%
\bibitem [{\citenamefont {Joshi}\ and\ \citenamefont
  {Mahesh}(2022)}]{Joshi2022}%
  \BibitemOpen
  \bibfield  {author} {\bibinfo {author} {\bibfnamefont {J.}~\bibnamefont
  {Joshi}}\ and\ \bibinfo {author} {\bibfnamefont {T.~S.}\ \bibnamefont
  {Mahesh}},\ }\bibfield  {title} {\bibinfo {title} {Experimental investigation
  of a quantum battery using star-topology nmr spin systems},\ }\bibfield
  {journal} {\bibinfo  {journal} {Physical Review A}\ }\textbf {\bibinfo
  {volume} {106}},\ \href {https://doi.org/10.1103/physreva.106.042601}
  {10.1103/physreva.106.042601} (\bibinfo {year} {2022})\BibitemShut {NoStop}%
\bibitem [{\citenamefont {Wootters}\ and\ \citenamefont
  {Zurek}(1982)}]{Wootters1982}%
  \BibitemOpen
  \bibfield  {author} {\bibinfo {author} {\bibfnamefont {W.~K.}\ \bibnamefont
  {Wootters}}\ and\ \bibinfo {author} {\bibfnamefont {W.~H.}\ \bibnamefont
  {Zurek}},\ }\bibfield  {title} {\bibinfo {title} {A single quantum cannot be
  cloned},\ }\href {https://doi.org/10.1038/299802a0} {\bibfield  {journal}
  {\bibinfo  {journal} {Nature}\ }\textbf {\bibinfo {volume} {299}},\ \bibinfo
  {pages} {802–803} (\bibinfo {year} {1982})}\BibitemShut {NoStop}%
\bibitem [{\citenamefont {Santos}\ \emph {et~al.}(2019)\citenamefont {Santos},
  \citenamefont {undefinedakmak}, \citenamefont {Campbell},\ and\ \citenamefont
  {Zinner}}]{Santos2019}%
  \BibitemOpen
  \bibfield  {author} {\bibinfo {author} {\bibfnamefont {A.~C.}\ \bibnamefont
  {Santos}}, \bibinfo {author} {\bibfnamefont {B.}~\bibnamefont
  {undefinedakmak}}, \bibinfo {author} {\bibfnamefont {S.}~\bibnamefont
  {Campbell}},\ and\ \bibinfo {author} {\bibfnamefont {N.~T.}\ \bibnamefont
  {Zinner}},\ }\bibfield  {title} {\bibinfo {title} {Stable adiabatic quantum
  batteries},\ }\bibfield  {journal} {\bibinfo  {journal} {Physical Review E}\
  }\textbf {\bibinfo {volume} {100}},\ \href
  {https://doi.org/10.1103/physreve.100.032107} {10.1103/physreve.100.032107}
  (\bibinfo {year} {2019})\BibitemShut {NoStop}%
\bibitem [{\citenamefont {{\AA}berg}(2014)}]{AbergCatalytic2014}%
  \BibitemOpen
  \bibfield  {author} {\bibinfo {author} {\bibfnamefont {J.}~\bibnamefont
  {{\AA}berg}},\ }\bibfield  {title} {\bibinfo {title} {Catalytic
  {{Coherence}}},\ }\href {https://doi.org/10.1103/PhysRevLett.113.150402}
  {\bibfield  {journal} {\bibinfo  {journal} {Physical Review Letters}\
  }\textbf {\bibinfo {volume} {113}},\ \bibinfo {pages} {150402} (\bibinfo
  {year} {2014})},\ \bibinfo {note}
  {\url{https://link.aps.org/doi/10.1103/PhysRevLett.113.150402}}\BibitemShut
  {NoStop}%
\bibitem [{\citenamefont {Kadian}\ \emph {et~al.}(2021)\citenamefont {Kadian},
  \citenamefont {Garhwal},\ and\ \citenamefont {Kumar}}]{KadianQuantum2021}%
  \BibitemOpen
  \bibfield  {author} {\bibinfo {author} {\bibfnamefont {K.}~\bibnamefont
  {Kadian}}, \bibinfo {author} {\bibfnamefont {S.}~\bibnamefont {Garhwal}},\
  and\ \bibinfo {author} {\bibfnamefont {A.}~\bibnamefont {Kumar}},\ }\bibfield
   {title} {\bibinfo {title} {Quantum walk and its application domains: {{A}}
  systematic review},\ }\href {https://doi.org/10.1016/j.cosrev.2021.100419}
  {\bibfield  {journal} {\bibinfo  {journal} {Computer Science Review}\
  }\textbf {\bibinfo {volume} {41}},\ \bibinfo {pages} {100419} (\bibinfo
  {year} {2021})},\ \bibinfo {note}
  {\url{https://www.sciencedirect.com/science/article/pii/S1574013721000599}}\BibitemShut
  {NoStop}%
\bibitem [{\citenamefont {Gross}\ \emph {et~al.}(2012)\citenamefont {Gross},
  \citenamefont {Nesme}, \citenamefont {Vogts},\ and\ \citenamefont
  {Werner}}]{GrossIndex2012}%
  \BibitemOpen
  \bibfield  {author} {\bibinfo {author} {\bibfnamefont {D.}~\bibnamefont
  {Gross}}, \bibinfo {author} {\bibfnamefont {V.}~\bibnamefont {Nesme}},
  \bibinfo {author} {\bibfnamefont {H.}~\bibnamefont {Vogts}},\ and\ \bibinfo
  {author} {\bibfnamefont {R.~F.}\ \bibnamefont {Werner}},\ }\bibfield  {title}
  {\bibinfo {title} {Index {{Theory}} of {{One Dimensional Quantum Walks}} and
  {{Cellular Automata}}},\ }\href {https://doi.org/10.1007/s00220-012-1423-1}
  {\bibfield  {journal} {\bibinfo  {journal} {Communications in Mathematical
  Physics}\ }\textbf {\bibinfo {volume} {310}},\ \bibinfo {pages} {419}
  (\bibinfo {year} {2012})}\BibitemShut {NoStop}%
\bibitem [{\citenamefont {Liu}\ \emph {et~al.}(2023)\citenamefont {Liu},
  \citenamefont {Culver}, \citenamefont {Harper},\ and\ \citenamefont
  {Roy}}]{LiuClassification2023}%
  \BibitemOpen
  \bibfield  {author} {\bibinfo {author} {\bibfnamefont {X.}~\bibnamefont
  {Liu}}, \bibinfo {author} {\bibfnamefont {A.~B.}\ \bibnamefont {Culver}},
  \bibinfo {author} {\bibfnamefont {F.}~\bibnamefont {Harper}},\ and\ \bibinfo
  {author} {\bibfnamefont {R.}~\bibnamefont {Roy}},\ }\href
  {https://doi.org/10.48550/arXiv.2308.02728} {\bibinfo {title} {Classification
  of {{Unitary Operators}} by {{Local Generatability}}}} (\bibinfo {year}
  {2023}),\ \Eprint {https://arxiv.org/abs/2308.02728} {arXiv:2308.02728
  [cond-mat, physics:quant-ph]} \BibitemShut {NoStop}%
\bibitem [{\citenamefont {Kitagawa}\ \emph {et~al.}(2010)\citenamefont
  {Kitagawa}, \citenamefont {Rudner}, \citenamefont {Berg},\ and\ \citenamefont
  {Demler}}]{KitagawaExploring2010}%
  \BibitemOpen
  \bibfield  {author} {\bibinfo {author} {\bibfnamefont {T.}~\bibnamefont
  {Kitagawa}}, \bibinfo {author} {\bibfnamefont {M.~S.}\ \bibnamefont
  {Rudner}}, \bibinfo {author} {\bibfnamefont {E.}~\bibnamefont {Berg}},\ and\
  \bibinfo {author} {\bibfnamefont {E.}~\bibnamefont {Demler}},\ }\bibfield
  {title} {\bibinfo {title} {Exploring topological phases with quantum walks},\
  }\href {https://doi.org/10.1103/PhysRevA.82.033429} {\bibfield  {journal}
  {\bibinfo  {journal} {Physical Review A}\ }\textbf {\bibinfo {volume} {82}},\
  \bibinfo {pages} {033429} (\bibinfo {year} {2010})},\ \bibinfo {note}
  {\url{https://link.aps.org/doi/10.1103/PhysRevA.82.033429}}\BibitemShut
  {NoStop}%
\bibitem [{\citenamefont {Liu}\ \emph {et~al.}(2018)\citenamefont {Liu},
  \citenamefont {Harper},\ and\ \citenamefont {Roy}}]{LiuChiral2018}%
  \BibitemOpen
  \bibfield  {author} {\bibinfo {author} {\bibfnamefont {X.}~\bibnamefont
  {Liu}}, \bibinfo {author} {\bibfnamefont {F.}~\bibnamefont {Harper}},\ and\
  \bibinfo {author} {\bibfnamefont {R.}~\bibnamefont {Roy}},\ }\bibfield
  {title} {\bibinfo {title} {Chiral flow in one-dimensional {{Floquet}}
  topological insulators},\ }\href {https://doi.org/10.1103/PhysRevB.98.165116}
  {\bibfield  {journal} {\bibinfo  {journal} {Physical Review B}\ }\textbf
  {\bibinfo {volume} {98}},\ \bibinfo {pages} {165116} (\bibinfo {year}
  {2018})}\BibitemShut {NoStop}%
\bibitem [{\citenamefont {Campaioli}\ \emph {et~al.}(2024)\citenamefont
  {Campaioli}, \citenamefont {Gherardini}, \citenamefont {Quach}, \citenamefont
  {Polini},\ and\ \citenamefont {Andolina}}]{CampaioliColloquium2024}%
  \BibitemOpen
  \bibfield  {author} {\bibinfo {author} {\bibfnamefont {F.}~\bibnamefont
  {Campaioli}}, \bibinfo {author} {\bibfnamefont {S.}~\bibnamefont
  {Gherardini}}, \bibinfo {author} {\bibfnamefont {J.~Q.}\ \bibnamefont
  {Quach}}, \bibinfo {author} {\bibfnamefont {M.}~\bibnamefont {Polini}},\ and\
  \bibinfo {author} {\bibfnamefont {G.~M.}\ \bibnamefont {Andolina}},\
  }\bibfield  {title} {\bibinfo {title} {Colloquium: {{Quantum}} batteries},\
  }\href {https://doi.org/10.1103/RevModPhys.96.031001} {\bibfield  {journal}
  {\bibinfo  {journal} {Reviews of Modern Physics}\ }\textbf {\bibinfo {volume}
  {96}},\ \bibinfo {pages} {031001} (\bibinfo {year} {2024})},\ \bibinfo {note}
  {\url{https://link.aps.org/doi/10.1103/RevModPhys.96.031001}}\BibitemShut
  {NoStop}%
\bibitem [{\citenamefont {Breuer}\ and\ \citenamefont
  {Petruccione}(2007)}]{Breuer2007}%
  \BibitemOpen
  \bibfield  {author} {\bibinfo {author} {\bibfnamefont {H.-P.}\ \bibnamefont
  {Breuer}}\ and\ \bibinfo {author} {\bibfnamefont {F.}~\bibnamefont
  {Petruccione}},\ }\href
  {https://doi.org/10.1093/acprof:oso/9780199213900.001.0001} {\emph {\bibinfo
  {title} {The Theory of Open Quantum Systems}}}\ (\bibinfo  {publisher}
  {Oxford University PressOxford},\ \bibinfo {year} {2007})\BibitemShut
  {NoStop}%
\bibitem [{\citenamefont {Allahverdyan}\ \emph {et~al.}(2004)\citenamefont
  {Allahverdyan}, \citenamefont {Balian},\ and\ \citenamefont
  {Nieuwenhuizen}}]{Nieuwenhuizen}%
  \BibitemOpen
  \bibfield  {author} {\bibinfo {author} {\bibfnamefont {A.~E.}\ \bibnamefont
  {Allahverdyan}}, \bibinfo {author} {\bibfnamefont {R.}~\bibnamefont
  {Balian}},\ and\ \bibinfo {author} {\bibfnamefont {T.~M.}\ \bibnamefont
  {Nieuwenhuizen}},\ }\bibfield  {title} {\bibinfo {title} {Maximal work
  extraction from finite quantum systems},\ }\href
  {https://doi.org/10.1209/epl/i2004-10101-2} {\bibfield  {journal} {\bibinfo
  {journal} {EPL}\ }\textbf {\bibinfo {volume} {67}},\ \bibinfo {pages} {565}
  (\bibinfo {year} {2004})}\BibitemShut {NoStop}%
\bibitem [{\citenamefont {Lenard}(1978)}]{LenardThermodynamical1978}%
  \BibitemOpen
  \bibfield  {author} {\bibinfo {author} {\bibfnamefont {A.}~\bibnamefont
  {Lenard}},\ }\bibfield  {title} {\bibinfo {title} {Thermodynamical proof of
  the {{Gibbs}} formula for elementary quantum systems},\ }\href
  {https://doi.org/10.1007/BF01011769} {\bibfield  {journal} {\bibinfo
  {journal} {Journal of Statistical Physics}\ }\textbf {\bibinfo {volume}
  {19}},\ \bibinfo {pages} {575} (\bibinfo {year} {1978})},\ \bibinfo {note}
  {\url{https://doi.org/10.1007/BF01011769}}\BibitemShut {NoStop}%
\bibitem [{\citenamefont {Pusz}\ and\ \citenamefont
  {Woronowicz}(1978)}]{PuszPassive1978}%
  \BibitemOpen
  \bibfield  {author} {\bibinfo {author} {\bibfnamefont {W.}~\bibnamefont
  {Pusz}}\ and\ \bibinfo {author} {\bibfnamefont {S.~L.}\ \bibnamefont
  {Woronowicz}},\ }\bibfield  {title} {\bibinfo {title} {Passive states and
  {{KMS}} states for general quantum systems},\ }\href
  {https://doi.org/10.1007/BF01614224} {\bibfield  {journal} {\bibinfo
  {journal} {Communications in Mathematical Physics}\ }\textbf {\bibinfo
  {volume} {58}},\ \bibinfo {pages} {273} (\bibinfo {year} {1978})},\ \bibinfo
  {note} {\url{https://doi.org/10.1007/BF01614224}}\BibitemShut {NoStop}%
\bibitem [{\citenamefont {Salvia}\ and\ \citenamefont
  {Giovannetti}(2021)}]{Salviadistribution2021}%
  \BibitemOpen
  \bibfield  {author} {\bibinfo {author} {\bibfnamefont {R.}~\bibnamefont
  {Salvia}}\ and\ \bibinfo {author} {\bibfnamefont {V.}~\bibnamefont
  {Giovannetti}},\ }\bibfield  {title} {\bibinfo {title} {On the distribution
  of the mean energy in the unitary orbit of quantum states},\ }\href
  {https://doi.org/10.22331/q-2021-08-03-514} {\bibfield  {journal} {\bibinfo
  {journal} {Quantum}\ }\textbf {\bibinfo {volume} {5}},\ \bibinfo {pages}
  {514} (\bibinfo {year} {2021})},\ \bibinfo {note}
  {\url{https://quantum-journal.org/papers/q-2021-08-03-514/}}\BibitemShut
  {NoStop}%
\bibitem [{Note1()}]{Note1}%
  \BibitemOpen
  \bibinfo {note} {While this equation also appeared in Ref.~\cite
  {Caravelli2020}, its implications were not fully appreciated
  therein.}\BibitemShut {Stop}%
\bibitem [{\citenamefont {Skrzypczyk}\ \emph {et~al.}(2013)\citenamefont
  {Skrzypczyk}, \citenamefont {Short},\ and\ \citenamefont
  {Popescu}}]{SkrzypczykExtracting2013}%
  \BibitemOpen
  \bibfield  {author} {\bibinfo {author} {\bibfnamefont {P.}~\bibnamefont
  {Skrzypczyk}}, \bibinfo {author} {\bibfnamefont {A.~J.}\ \bibnamefont
  {Short}},\ and\ \bibinfo {author} {\bibfnamefont {S.}~\bibnamefont
  {Popescu}},\ }\href@noop {} {\bibinfo {title} {Extracting work from quantum
  systems}},\ \bibinfo {howpublished} {\url{http://arxiv.org/abs/1302.2811}}
  (\bibinfo {year} {2013}),\ \Eprint {https://arxiv.org/abs/1302.2811}
  {arXiv:1302.2811 [quant-ph]} \BibitemShut {NoStop}%
\bibitem [{\citenamefont {Skrzypczyk}\ \emph {et~al.}(2014)\citenamefont
  {Skrzypczyk}, \citenamefont {Short},\ and\ \citenamefont
  {Popescu}}]{SkrzypczykWork2014}%
  \BibitemOpen
  \bibfield  {author} {\bibinfo {author} {\bibfnamefont {P.}~\bibnamefont
  {Skrzypczyk}}, \bibinfo {author} {\bibfnamefont {A.~J.}\ \bibnamefont
  {Short}},\ and\ \bibinfo {author} {\bibfnamefont {S.}~\bibnamefont
  {Popescu}},\ }\bibfield  {title} {\bibinfo {title} {Work extraction and
  thermodynamics for individual quantum systems},\ }\href
  {https://doi.org/10.1038/ncomms5185} {\bibfield  {journal} {\bibinfo
  {journal} {Nature Communications}\ }\textbf {\bibinfo {volume} {5}},\
  \bibinfo {pages} {4185} (\bibinfo {year} {2014})},\ \bibinfo {note}
  {\url{https://www.nature.com/articles/ncomms5185}}\BibitemShut {NoStop}%
\bibitem [{\citenamefont {{\L}obejko}(2021)}]{Lobejkotight2021}%
  \BibitemOpen
  \bibfield  {author} {\bibinfo {author} {\bibfnamefont {M.}~\bibnamefont
  {{\L}obejko}},\ }\bibfield  {title} {\bibinfo {title} {The tight {{Second
  Law}} inequality for coherent quantum systems and finite-size heat baths},\
  }\href {https://doi.org/10.1038/s41467-021-21140-4} {\bibfield  {journal}
  {\bibinfo  {journal} {Nature Communications}\ }\textbf {\bibinfo {volume}
  {12}},\ \bibinfo {pages} {918} (\bibinfo {year} {2021})},\ \bibinfo {note}
  {\url{https://www.nature.com/articles/s41467-021-21140-4}}\BibitemShut
  {NoStop}%
\bibitem [{Note2()}]{Note2}%
  \BibitemOpen
  \bibinfo {note} {The double-sided energy ladder itself might be regarded as
  being unphysical because it lacks a ground state. However, we show that even
  if such a spectrum were assumed to exist, a UC protocol can still not be
  generated.}\BibitemShut {Stop}%
\bibitem [{\citenamefont {Kitaev}(2006)}]{KitaevAnyons2006}%
  \BibitemOpen
  \bibfield  {author} {\bibinfo {author} {\bibfnamefont {A.}~\bibnamefont
  {Kitaev}},\ }\bibfield  {title} {\bibinfo {title} {Anyons in an exactly
  solved model and beyond},\ }\href {https://doi.org/10.1016/j.aop.2005.10.005}
  {\bibfield  {journal} {\bibinfo  {journal} {Annals of Physics}\ }\textbf
  {\bibinfo {volume} {321}},\ \bibinfo {pages} {2} (\bibinfo {year}
  {2006})}\BibitemShut {NoStop}%
\bibitem [{\citenamefont {Caravelli}\ \emph {et~al.}(2020)\citenamefont
  {Caravelli}, \citenamefont {Coulter-De~Wit}, \citenamefont {García-Pintos},\
  and\ \citenamefont {Hamma}}]{Caravelli2020}%
  \BibitemOpen
  \bibfield  {author} {\bibinfo {author} {\bibfnamefont {F.}~\bibnamefont
  {Caravelli}}, \bibinfo {author} {\bibfnamefont {G.}~\bibnamefont
  {Coulter-De~Wit}}, \bibinfo {author} {\bibfnamefont {L.~P.}\ \bibnamefont
  {García-Pintos}},\ and\ \bibinfo {author} {\bibfnamefont {A.}~\bibnamefont
  {Hamma}},\ }\bibfield  {title} {\bibinfo {title} {Random quantum batteries},\
  }\bibfield  {journal} {\bibinfo  {journal} {Physical Review Research}\
  }\textbf {\bibinfo {volume} {2}},\ \href
  {https://doi.org/10.1103/physrevresearch.2.023095}
  {10.1103/physrevresearch.2.023095} (\bibinfo {year} {2020})\BibitemShut
  {NoStop}%
\end{thebibliography}
%apsrev4-2.bst 2019-01-14 (MD) hand-edited version of apsrev4-1.bst
%Control: key (0)
%Control: author (8) initials jnrlst
%Control: editor formatted (1) identically to author
%Control: production of article title (0) allowed
%Control: page (0) single
%Control: year (1) truncated
%Control: production of eprint (0) enabled
%

%%%%%%%%%%%%%%%%%%% APPENDIX / SUPPLEMENTARY MATERIAL
\clearpage
\newpage

% %%%%%%%%%% Prefix a "S" to all equations, figures, tables and reset the counter %%%%%%%%%%
\setcounter{equation}{0}
\setcounter{figure}{0}
\setcounter{section}{0}
\setcounter{table}{0}
\setcounter{page}{1}
\makeatletter
\renewcommand{\theequation}{S\arabic{equation}}
\renewcommand{\thepage}{S-\arabic{page}}
\renewcommand{\thesection}{S\arabic{section}}

\begin{widetext}

\section*{Supplemental Material to ``Universally-Charging Protocols for Quantum Batteries: A No-Go Theorem''}

\section{Derivation of UC protocol for the auxiliary qubit model} \label{app:aux_qubit}
Here, we derive the UC protocol corresponding to the auxiliary qubit model given by
\begin{align}
    H(t) = \begin{cases}
        -\pi \left( \eye \otimes \sigma_x \right)&  0 \leq t < \frac{1}{2}, \\
        {\pi} \left(s^\dagger \otimes \sigma_+ + s \otimes \sigma_- \right) &  \frac{1}{2} \leq t \leq 1. \\
    \end{cases} \label{eq:protocol_supp}
\end{align}
In order to understand the behavior of $U$, it is sufficient to focus on the effects of $U$ on the basis $\mathcal B = \{\ket{k,a} | k=1,\dotsc, N; a=\uparrow, \downarrow \}$ of $\h_\text{total}$.
First, we note that 
\bseq
\begin{align}
    U \left( 0 \rightarrow 1/2 \right) &= \exp \left(i  \frac{\pi}{2}  (\eye \otimes \sigma^x )\right) \\
    &= \eye \otimes e^{i\frac{\pi}{2} \sigma^x} = i \ \eye \otimes \sigma^x.
\end{align}
\eseq
Similarly, it can be seen that for $1/2 \leq t \leq 1$, $H(t)$ is block diagonal in $\mathcal{B}$ basis. Furthermore, for all $n\in \{ 1,\dotsc, N\}$, $H$ maps $\ket{n,\downarrow}$ to $\ket{n+1,\uparrow}$ and vice versa.
Therefore, we have
\begin{align*}
    U \left( 1 / 2 \rightarrow 1 \right) \ket{n,\downarrow} &= -i \ket{n+1,\uparrow} \quad \text{if } n\in \{1,\dotsc N-1\} \\
    U \left( 1 / 2 \rightarrow 1 \right) \ket{N,\downarrow} &= \ket{N,\downarrow} \\
    U \left( 1 / 2 \rightarrow 1 \right) \ket{n,\uparrow} &= -i \ket{n-1, \downarrow} \quad \text{if } n \in \{ 2,\dotsc, N\} \\
    \text{and }U \left( 1 / 2 \rightarrow 1 \right) \ket{1,\uparrow} &= \ket{1,\uparrow}.
\end{align*}

Thus, the time evolution of the full protocol is given by 
\bseq
\begin{align} \label{eq:unitary_protocol_ancilla_finite_dim}
    U &\equiv U(1 / 2 \rightarrow 1) U(0 \rightarrow 1 / 2),\\ 
    \text{with }    U \ket{n,\uparrow} &= \begin{cases}
        \ket{n+1, \uparrow} & \text{if } n \in \{ 1,\dotsc, N-1\}\\
        i \ket{N,\downarrow} & \text{if } n = N.
    \end{cases}\\
    \text{and } U \ket{n,\downarrow} &= \begin{cases}
        \ket{n-1,\downarrow } & \text{if }n \in \{2,\dotsc, N\} \\
        i\ket{1, \uparrow} & \text{if }n = 1.
    \end{cases}
\end{align}
\eseq

The action of the protocol on the battery Hilbert space $\h_B$, when the control qubit is set to $\ket \uparrow$ is then found to be
\bseq
\begin{align}
    \map_{\ket{\uparrow}} [\rho] &\coloneqq \Tr_E [U (\rho_B \otimes \ket{\uparrow}\bra{\uparrow}) U^\dagger] \\
    &= s^\dagger \rho s + \Pi_{N} \rho \Pi_{N},
\end{align}
\eseq
with $\Pi_N = \ket N \bra N$.
This protocol is therefore universally charging, since it raised the energy of all states positively, except for the highest-energy state, whose energy is not changed.

When the control qubit is set to $\ket \downarrow$, the protocol in universally discharging, since 
\begin{align}
    \map_{\ket \downarrow} [\rho]  &\coloneqq \Tr_E [U (\rho_B \otimes \ket{\downarrow}\bra{\downarrow}) U^\dagger] \\
    &= s \rho s^\dagger + \Pi_{1} \rho \Pi_{1}, 
\end{align}
where $\Pi_1 = \ket 1 \bra 1$.

\section{Haar averages for CPTP maps} \label{app:Haar_CPTP}
As seen before [see Eq.~\eqref{eq:Haar_average_over_density_matrices}], the Haar average of $\Delta E$ over rotations of any density matrix is zero for a unitary protocol. 
For arbitrary CPTP maps, the average values are generically non-zero.
To see this, we consider the Kraus decomposition of a map $\map$, expressed as $\map[\rho] = \sum_k L_k \rho L_k^\dagger$.
Then, using the formula $\ov{GAG^\dagger}^G=\frac{\tr{A}}{\dim \h_B} \eye$, the Haar-averaged energy is found to be
\bseq
\begin{align}
    \ov{\Delta E} &= \ov{\Tr [H_B (\map [G \rho G^\dagger ] - G \rho G^\dagger)]} \\
    &= \ov {\Tr[H_B (\sum_k L_k G \rho G^\dagger L_k^\dagger  - G \rho G^\dagger) ]} \\
    &= \frac{1}{\dim \h_B}\Tr [H_B (\sum_k L_k (\Tr \rho) \eye  L_k^\dagger - \Tr \rho \eye )] \\
    &= \frac{1}{\dim \h_B}\Tr [H_B (\sum_k L_k L_k^\dagger - \eye)].
\end{align}
\eseq
We note that the RHS is generically non-zero, since even though $\sum_k L_k^\dagger L_k = \eye$, $\sum_k L_k L_k^\dagger \neq \eye$ in general, unless the map is non-unitary.

For the UC protocol~\eqref{eq:map_uparrow} discussed above, the Haar averaged $\Delta E$ evaluates to
\bseq
\begin{align}
    \ov{\Delta E} &= \Tr [H_B (s^\dagger s  + \Pi_N - \eye)] \\
    &= \Tr [H_B (\{\sum_{k=1}^N \Pi_k \} + \Pi_N - \eye)] \\
    &= \Tr [H_B (\Pi_N - \Pi_1)] \\
    &= E_N - E_1 > 0.
\end{align}
\eseq
\end{widetext}

\end{document}